\documentclass[11pt]{article}

\usepackage{amsmath}
\usepackage{amssymb}
\usepackage{amsthm}
\usepackage{latexsym}
\usepackage{color}
\usepackage{graphicx}
\usepackage{appendix}
\usepackage{color} 
\usepackage{enumerate}
\usepackage[T1]{fontenc}
\usepackage[english]{babel}
\usepackage[table]{xcolor}
\usepackage{url}
\usepackage{geometry}
\DeclareSymbolFont{calletters}{OMS}{cmsy}{m}{n}
\DeclareSymbolFontAlphabet{\mathcal}{calletters}

\def\be{\begin{eqnarray}}
\def\ee{\end{eqnarray}}

\def\b*{\begin{eqnarray*}}
\def\e*{\end{eqnarray*}}

\newtheorem{Theorem}{Theorem}[part]
\newtheorem{Definition}{Definition}[part]
\newtheorem{Proposition}{Proposition}[part]

\newtheorem{Assumption}{Assumption}[part]
\newtheorem{Lemma}{Lemma}[part]

\newtheorem{Remark}{Remark}[part]

\makeatletter \@addtoreset{equation}{section}

\@addtoreset{Definition}{section}

\@addtoreset{Theorem}{section}

\@addtoreset{Proposition}{section}

\@addtoreset{Property}{section}

\@addtoreset{Assumption}{section}

\@addtoreset{Corollary}{section}

\@addtoreset{Lemma}{section}

\@addtoreset{Remark}{section}

\@addtoreset{Example}{section}

\@addtoreset{Condition}{section}

\newcommand{\No}[1]{\left\|#1\right\|}     
\newcommand{\abs}[1]{\left|#1\right|}     

\def \E{\mathbb{E}}
\def \F{\mathbb{F}}
\def \H{\mathbb{H}}

\def \P{\mathbb{P}}
\def \Q{\mathbb{Q}}
\def \R{\mathbb{R}}

\def\Cc{{\cal C}}

\def\Ec{{\cal E}}
\def\Fc{{\cal F}}

\def\Mc{{\cal M}}

\def\Tr#1{{\rm Tr}\left[#1\right]}

\def \Sum{\displaystyle\sum}
\def \Prod{\displaystyle\prod}

\def\esup{{\rm ess \, sup}}

\def\={\;=\;}
\def\.{\;.}

\def\eps{\varepsilon}

\def\reff#1{{\rm(\ref{#1})}}

\def\1{{\bf 1}}
\def \ep{\hbox{ }\hfill$\Box$}

\def \proof{{\noindent \bf Proof. }}
\def\eps{\epsilon}

\def\b*{\begin{eqnarray*}}
\def\e*{\end{eqnarray*}}

 \def\normeL2#1{\left\|{#1}\right\|_{L^2}}
 
 \title{Contracting theory with competitive interacting Agents}

 \author{Romuald {\sc Elie} \thanks{Universit\'e Paris-Est Marne-la-Vall\'ee  $\&$ Projet MathRisk INRIA, romuald.elie@univ-mlv.fr. Research partially supported by the ANR grant LIQUIRISK and the Chair Finance and Sustainable Development.}\and Dylan {\sc Possama\"{i}} \footnote{CEREMADE, Universit\'e Paris Dauphine, possamai@ceremade.dauphine.fr.}}
 
             \date{\today}

\begin{document}

\maketitle

\begin{abstract}
In a framework close to the one developed by Holmstr\"om and Milgrom \cite{hm}, we study the optimal contracting scheme between a Principal and several Agents. Each hired Agent is in charge of one project, and can make efforts towards managing his own project, as well as impact (positively or negatively) the projects of the other Agents. Considering economic agents in competition with relative performance concerns, we derive the optimal contracts in both first best and moral hazard settings. The enhanced resolution methodology relies heavily on the connection between Nash equilibria and multidimensional quadratic BSDEs. The optimal contracts are linear and each agent is paid a fixed proportion of the terminal value of all the projects of the firm. Besides, each Agent receives his reservation utility, and those with high competitive appetence are assigned less volatile projects, and shall even receive help from the other Agents. From the principal point of view, it is in the firm interest in our model to strongly diversify the competitive appetence of the Agents.
\end{abstract}

\vspace{1em} 

\vspace{1em}
\noindent
{\noindent \textit{AMS 2010 subject classification:} 91B40,  91A15, 93E20\\[1mm]
\normalsize
}
{\noindent \textit{JEL classification:} C61, C73, D82, J33, M52
\normalsize
}

\section{Introduction} 
\label{sec:introduction} 
By and large, it has now become common knowledge among the economists, that almost everything in economics was to a certain degree a matter of incentives: incentives to work hard, to produce, to study, to invest, to consume reasonably... Starting from the 70s, the theory of contracts evolved from this acknowledgment and the fact that such situations could not be reproduced using the general equilibrium theory. In the corresponding typical situation, a Principal (who takes the initiative of the contract) is (potentially) imperfectly informed about the actions of an Agent (who accepts or rejects the contract). The goal is to design a contract that maximizes the utility of the Principal while that of the Agent is held to a given level. Of course, the form of the optimal contracts typically depends on whether these actions are observable/contractible or not, and on whether there are characteristics of the Agent that are unknown to the Principal. These problems are fundamentally linked to designing optimal incentives, and are therefore present in a very large number of situations (see Bolton and Dewatripont \cite{BD} or Laffont and Martimort \cite{laffont} for many examples). 

\vspace{0.5em}
The easiest problem corresponds to the case were the Principal is actually perfectly informed about the actions of the Agent, and just has to find a way to optimally share the risks associated to the project he has hired the Agent for, between the two of them: this is the so-called risk-sharing problem or first-best. Early studies of the risk-sharing problem can be found, among others, in Borch \cite{borch}, Wilson \cite{wilson} or Ross \cite{ross}. Since then, a large literature has emerged, solving very general risk-sharing problems, for instance in a framework with several Agents and recursive utilities (see Duffie et al. \cite{duffie}), or for studying optimal compensation of portfolio managers (see Ou-Yang \cite{ou} or Cadenillas et al. \cite{CCZ}). 

\vspace{0.5em}
A more complicated situation arises in the so-called moral hazard case, or second-best, where the Principal cannot observe (or contract upon) the actions chosen by the Agent. For a long time, these problems were only considered in discrete-time or static settings\footnote{For very early moral hazard models, introducing the so-called first-order approach and then later its rigorous justification, see Zeckhauser \cite{zec}, Spence and Zeckhauser \cite{szec}, or Mirrlees \cite{mir1,mir3,mir4}, as well as the seminal papers by Grossman and Hart \cite{gross}, Jewitt, \cite{jew}, Holmstr\"om \cite{holm} or Rogerson \cite{roger}.}, which are in general quite hard to solve, and one had to wait for the seminal paper by Holmstr\"om and Milgrom \cite{hm} to witness the treatment of specific moral hazard problems in a continuous time framework. Their work was generalized by Sch\"attler and
Sung \cite{SS3,SS7}, Sung \cite{Su, S2},
M\"uller \cite{M1,Mul2},
and Hellwig and Schmidt \cite{HS8}, using a dynamic programming and martingales approach, which is classical in stochastic control theory (see also the survey paper by Sung \cite{S3} for more references). This approach has then been extended to a very general framework in recent works by Cvitani\'c, Possama\"i and Touzi \cite{cpt,cpt2}\footnote{Another, and in some cases more general, approach is to use the so-called stochastic maximum principle and FBSDEs to characterize the optimal
compensation. This is the strategy used by Williams \cite{Williams} and
Cvitani\'c, Wan and Zhang \cite{CWZ5, CZ2}, and more recently, by Djehiche and Hegelsson \cite{dje1,dje2}. We also refer the reader to the excellent monograph of Cvitani\'c and Zhang \cite{CvitanicZhang} for a systematic presentation of this approach.}. Yet another recent seminal paper is the one by Sannikov \cite{sann}, who
finds a tractable model with a random time of
retiring the Agent and with continuous payments, rather than a lump-sum payment at the terminal time. Since then, a growing literature extending the above models has emerged, see the illuminating survey paper \cite{sann2} for a quite comprehensive list of references.

\vspace{0.5em}
Another possible extension of the moral hazard problem lies in considering that the Principal no longer hires just one Agent, but several of them, to manage one or several projects on his behalf, while being able to interact with each other. This is the so-called multi-Agent problem. Early works in that direction, again in one-period frameworks, include Holmstr\"om \cite{holmm}, Mookherjee \cite{mook}, Green and Stokey \cite{gs}, Harris et al. \cite{harr}, Nalebuff and Stiglitz \cite{ns} or Demski and Sappington \cite{ds}. As far as we know, the first extension to continuous time is due to Koo et al. \cite{koo}, who considered roughly the same model as Holmstr\"om \cite{holmm}, by using the martingale approach described above. Of course, as soon as one starts to consider contracting situation involving several agents, the question of these agents comparing themselves to each other becomes quite relevant. Hence, contemporary to the latter studies, several researchers tried to understand the impact of the so-called inequity aversion, as formulated by Fehr and Schmidt \cite{fehr}, on agency costs. The idea is that agents working in a firm dislike inequity, in the sense that an agent suffers a utility loss if another agent conducting a similar task receives a higher wage. The first papers which examined moral hazard problems in view of inequity aversion are Englmaier and Wambach \cite{ew,ew2}, where a single risk-averse Agent compares himself to the Principal, as well as and Fehr et al. \cite{fehr2}. Later studies involving a Principal and several Agents include Itoh \cite{itoh}, Rey-Biel \cite{rey}, Bartling and von Siemens \cite{barts1,barts2}, or Grund and Sliwka \cite{grund}, which analyze the consequences of inequity aversion for team incentives. Demougin and Fluet \cite{demou1,demou2,demou3} as well as Neilson and Stowe \cite{neilson1,neilson2} or more recently Kragl \cite{kragl}, show that a trade-off between incentive effect and inequity costs also arises with individual performance pay. Another related paper is Dur and Glazer \cite{dg}, where the authors study optimal contracts when workers envy their boss. A continuous time extension of the problem has also been studied by Goukasian and Wan \cite{gw}, still in a context where Agents exhibit envy and jealousy towards their co-workers. Almost all these works show that envious behavior is destructive for organizations.

\vspace{0.5em}
Our paper still follows the strand of literature described above, but  slightly departs from it in the sense that we aim at studying the impact on optimal incentives and contracts of competitiveness among Agents in a firm. More precisely, we place ourselves in the continuous-time model of Holmstr\"om and Milgrom \cite{hm}, but we assume that the utility derived by the Agents is increasing in their wages, but also on their performance, compared to other Agents in the firm. As far as we know, such a setting has not been considered yet in the contracting theory literature, though our take on the problem is inspired by Espinosa and Touzi \cite{et}, which studied a classical problem of portfolio optimization in financial markets, where the utility of the investors depended also on how well they performed compared to other related investors in the market. In our model, one Principal hires many Agents to manage several, possibly correlated projects, on his behalf. Furthermore, given the intrinsic notion of competition, we assume that the Agents can decide to work for their own project, but also that they can try to either help or decrease the value of the projects managed by the other agents. Besides, all Agents have a Nash equilibrium type behavior, that is to say that they all compute their best reaction functions given a strategy chosen by the other Agents, and then agree on an equilibrium. We also assume that we are in a moral hazard setting, and that the Principal can only observe the outcomes of each project managed (event though we also solve the first-best problem in Section \ref{sec:first} to have a convenient benchmark setting). 

\vspace{0.5em}
From the mathematical point of view, we are therefore solving a Stackelberg game between the Principal and the Agents, the latter also playing a non-zero-sum game between each other. Since we work in continuous-time, we make a very important use of the theory of backward stochastic differential equations (BSDEs for short, see \cite{ekpq} for more details about the theory), which is the probabilistic and non-Markovian counterpart of the Hamilton-Jacobi Bellman semi-linear partial differential equation which characterizes the value function of the Agent, for a given Markovian contract. Thanks to this theory, we establish that existence of a Nash equilibrium for the Agents, is basically equivalent to finding a solution to a multidimensional BSDE whose non-linearity exhibits quadratic growth in the so-called control variable. The wellposedness theory for such equations is still not well understood at all (see \cite{tev,fdr,f,cn,khz,kp,kp2,ht,jam,lt} for more details, partial results and counterexamples), but we actually circumvent this problem by imposing wellposedness as a requirement for admissibility of the contracts. Such a restriction could be seen as too important, but the intuition is that if there is a contract for which there does not exist any Nash equilibrium between the Agents, then the Principal will never propose it, as he will never be able to check the incentive compatibility constraint, or in other words he would never know the amount of work the Agents will consent to provide. Anyway, we show in quite general situations that we can obtain the optimal compensation scheme explicitly, and that it is indeed admissible in the above sense. 

\vspace{0.5em}
Furthermore, we once again recover the seminal result of \cite{hm}, and prove that the optimal contract is a linear function of all the terminal values of the projects managed by the Agents. Therefore, to provide correct incentives in a framework where Agents compare their own performances, the Principal has to reward them using not only their own projects, but also the ones managed by the other Agents. More surprisingly, we also show that in a situation where one Agent is very competitive compared to another one, it may be more profitable for the Principal to give incentives to both Agents to work for the project of the very competitive Agent, but also to decrease the value of the project managed by the less competitive Agent. The intuition here is that if one Agent is very competitive, it is less costly for the Principal to make sure that he will be the best in terms of performances, than to compensate him with a higher salary, all the more since other Agents only care about their wages.

\vspace{0.5em}
Given such a result, one could argue that, as envy or jealousy, competitiveness among Agents is destructive for the firm. However, we show that, at least in our model, this is never the case. Indeed, we prove that given a choice between several type of Agents, with different appetences for competition, the Principal will either hire Agents with  the largest difference in their appetence, if their marginal costs of working are not too high, or with a fixed difference otherwise, but never identical Agents. Therefore, the Principal can always get a higher benefit from a diversity of competition profiles  among the firm. Furthermore, as in \cite{hm}, since all the Agents end up receiving their reservation utility, they enjoy the same resulting happiness whether they are competitive or not (assuming of course that their reservation utilities would stay the same), competition in our model is also socially better. This is in stark contrast with situations where Agents compare their salaries.

\vspace{0.5em}
The rest of the paper is organized as follows. We describe our general model in Section \ref{sec:model}. Then, Section \ref{sec:first} is devoted to solving the first-best problem in a very general setting, for which we obtain several explicit solutions in more specific contexts. In Section \ref{sec:second}, we move on to the second-best problem, for which we provide in a very general framework the HJB equations satisfied by the value function of the Principal, and which we solve explicitly in a context similar to the one of \cite{hm}.  

\vspace{2.5em}
{\bf Notations:} Let $\mathbb{N}^\star :=\mathbb{N}\setminus\{0\}$ and let $\mathbb{R}_+^\star $ be the set of real positive numbers. Throughout this paper, for every $p$-dimensional vector $b$ with $p\in \mathbb{N}^\star $, we denote by $b^{1},\ldots,b^{p}$ its coordinates, for any $1\leq i\leq p$, by $b^{-i}\in\R^{p-1}$ the vector obtained by suppressing the $i$th coordinate of $b$, and for $p>1$
 $$\bar b^{-i}:=\frac{1}{p-1}\sum_{j=1,\ j\neq i}^pb^j.$$
 For $\alpha,\beta \in \R^p$ we also denote by $\alpha\cdot \beta$ the usual inner product, with associated norm $\No{\cdot}$, which we simplify to $|\cdot|$ when $p$ is equal to $1$. We also let ${\bf 1}_p$ be the vector of size $p$ whose coordinates are all equal to $1$. For any $(l,c)\in\mathbb N^\star \times\mathbb N^\star $, $\mathcal M_{l,c}(\mathbb R)$ will denote the space of $l\times c$ matrices with real entries. Elements of the matrix $M\in\mathcal M_{l,c}$ will be denoted by $(M^{i,j})_{1\leq i\leq l,\ 1\leq j\leq c}$, and the transpose of $M$ will be denoted by $M^\top$. We identify $\mathcal M_{l,1}$ with $\R^l$. When $l=c$, we let $\mathcal M_{l}(\mathbb R):=\mathcal M_{l,l}(\mathbb R)$. For any $x\in\mathcal M_{l,c}(\mathbb R)$, and for any $1\leq i\leq l$ and $1\leq j\leq c$, $x^{i,:}\in \mathcal M_{1,c}(\R)$ and $x^{:,j}\in\R^l$ will denote respectively the $i$th row and the $j$th column of $M$. Moreover, for any $x\in\mathcal M_{l,c}(\mathbb R)$ and any $1\leq j\leq c$, $x^{:,-j}\in\mathcal M_{l,c-1}$ will denote the matrix $x$ without the $j$th column. For any $x\in\mathbb R^p$, ${\rm diag}(x)\in\mathcal M_{p}(\mathbb R)$ will stand for the matrix whose diagonal is $x$ and for which off-diagonal terms are $0$, and $I_p$ will be the identity matrix in $\mathcal M_p(\mathbb R)$. For any $x\in\mathcal M_{l,c}(\R)$ and any $y\in\R^l$, we also define, for any $i=1,\dots,c$, $y\otimes_ix\in\mathcal M_{l,c+1}(\R)$ as the matrix whose column $j=1,\dots,i-1$ is equal to the $j$th column of $x$, whose column $j=i+1,\dots,c+1$ is equal to the $(j-1)$th column of $x$, and whose $i$th column is $y$. We also abuse notations and denote for any $x\in \mathcal M_p(\R)$ by $\No{x}$ the operator norm of $x$ associated to the Euclidean norm on $\R^p$. The trace of a matrix $M\in\mathcal M_l(\R)$ will be denoted by $\Tr{M}$.
 
 \vspace{0.5em}
For any finite dimensional vector space $E$, with given norm $\No{\cdot}_E$, we also introduce the so-called Morse-Transue space on  a given probability space$(\Omega,\mathcal F,\P)$ (we refer the reader to the monographs \cite{rao1,rao2} for more details), defined by
\begin{equation}\label{eq:morse}
M^\phi(E):=\left\{\xi:=\Omega\longrightarrow E,\ \text{measurable},\ \mathbb E\left[\phi(a\xi)\right]<+\infty,\ \text{for any $a\geq 0$}\right\},
\end{equation}
where $\phi:E\longrightarrow\R$ is the Young function, i.e. $\phi:x\mapsto\exp(\No{x}_E)-1$. Then, if $M^\phi(E)$ is endowed with the norm
$\No{\xi}_{\phi}:=\inf\{k>0,\  \mathbb E\left[\phi\left(\xi/ k\right)\right]\leq 1\},$
is a (non-reflexive) Banach space.

\section{Formulation of the problem}\label{sec:model}

 This opening section sets up properly the problem of interest. We first define the dynamics of the firm. Second, we model the impact of the economic choices made by the system of Agent. We then define properly the set of admissible strategies for the Agents in such system, and we finally indicate the objective function of each Agent, as well as the one of the Principal.

\subsection{The firm dynamics}
We consider a model where a Principal wishes to hire $N\geq 1$ Agents, in order to take care of $N$ different projects. Each Agent, if hired, will have the responsibility of a risky project. In order to define precisely the outcome of the actions chosen by each Agent, let us start by fixing some notations.

\vspace{0.5em}
\noindent We fix a deterministic time horizon $T>0$. We work on a given probability space $(\Omega,\mathcal F,\P)$ carrying an $N$-dimensional Brownian motion $W$. Each component of $W$ will drive the noise associated to one project of the firm. We denote by $\F:=(\mathcal F_t)_{0\leq t\leq T}$ the (completed) natural filtration of $W$. As is customary in the Principal/Agent literature, we will work under the so-called weak formulation of the problem.

\vspace{0.5em}
\noindent To define this rigorously, let us start by defining the so-called output process $X$ of the firm, which is $\R^N$-valued,
\begin{equation}\label{eq:defX}
X_t:=\int_0^t\Sigma_sdW_s,\ 0\leq t\leq T, \ a.s.,
\end{equation}
where for any $t\in[0,T]$, $\Sigma_t\in\mathcal M_N(\R).$ Each component of the vector $X$ corresponds to one project of the firm and the matrix $\Sigma$ characterizes their correlation. Our assumption on $\Sigma$ is as follows
\begin{Assumption}\label{assump:sigma}
The map $\Sigma:[0,T]\longrightarrow \mathcal M_N(\R)$ is $($Borel$)$ measurable, bounded and such that for any $t\in[0,T]$, $\Sigma_t$ is invertible.
\end{Assumption}

\subsection{The impact of the effort provided by the system of Agents}

\vspace{0.5em}
We will consider that each Agent will be assigned a project but  can impact both his own project as well as the projects of the other Agents. Hence, the controls of all the Agents will be given by a matrix $a\in\mathcal M_N(\R)$, such that for any $1\leq i , j\leq N$, $a^{i,j}$ represents the action of Agent $j$ for the project managed by Agent $i$. In other words, each Agent can choose to make efforts towards managing his own project, but he can also decide to impact (positively or negatively) the projects of the other Agents.

\vspace{0.5em}
We also introduce for any $1\leq i\leq N$ the maps $b^i:[0,T]\times\R^N\times\R^N\longrightarrow \R$, which are assumed to satisfy the following
\begin{Assumption}\label{assump:b}
For every $i=1,\dots,N$, and every $(t,x)\in[0,T]\times\R^N$, the maps $a\longmapsto b^i(t,a,x)$ are $C^1$, and for every $a\in\R^N$, the maps $x\longmapsto b^i(t,a,x)$ are uniformly Lipschitz continuous. We also assume that for some some constant $C>0$,
\begin{equation}\label{eq:growthb}
\abs{b^i(t,a,x)}\leq C(1+\No{a}+\No{x}),\ \No{\nabla_a b^i(t,a,x)}\leq C,\ \text{for any $(t,a,x)\in[0,T]\times\R^N\times\R^N$}.
\end{equation}
\end{Assumption}
For any $(t,a,x)\in[0,T]\times \mathcal M_N(\R)\times\R^N$, we also denote by $b(t,a,x)$ the vector of $\R^N$ whose $i$th coordinate is $b^i(t,a^{:,i},x)$.

\vspace{0.5em}
Notice that by Assumption \ref{assump:b}, for any $\mathcal M_N(\R)$-valued and $\F$-progressively measurable process $a$, the following SDE
\begin{equation}\label{eq:strong}
Z_t=\int_0^tb(s,a_s,Z_s)ds+\int_0^t\Sigma_sdW_s, t\in[0,T],\ \P-a.s.,
\end{equation}
admits a unique strong solution denoted $X^a$.

\vspace{0.5em}
For any $\mathcal M_N(\R)$-valued $\F$-progressively measurable processes $a$ satisfying 
\begin{equation}\label{eq:dd}
\mathcal E\left(\int_0^Tb\left(s,a_s,X_s\right)\cdot \Sigma^{-1}_s dW_s\right)\ \text{has moments of order $(1+\eps)$ for some $\eps>0$,}
\end{equation}
 we can then define a new probability measure $\P^a$ on $(\Omega,\mathcal F)$, equivalent to $\P$, with
$$\frac{d\P^a}{d\P}:=\mathcal E\left(\int_0^Tb(s,a_s,X_s)\cdot \Sigma^{-1}_s dW_s\right).$$
By Girsanov's theorem, we then know that for any $a\in\mathcal A$, the following $\R^N$-valued process
$$W^a_t:=W_t-\int_0^t\Sigma^{-1}_s b(s,a_s,X_s)ds,\ 0\leq t\leq T,\ a.s.,$$
is a Brownian motion under $\P^a$, so much so that we can rewrite \reff{eq:defX}, for any $a\in\mathcal A$, as
$$X_t=\int_0^tb(s,a_s,X_s)ds+\int_0^t\Sigma_s dW^a_s,\ 0\leq t\leq T,\ a.s.$$
 Moreover, it is clear that $\P^a$ coincides with the probability measure $\widetilde \P^a:=\P\circ(X^{a(X^a_\cdot)})^{-1}$ obtained from the strong solution of \eqref{eq:strong}.

\subsection{The admissible efforts for the Agents}

\vspace{0.5em}
We introduce for any $1\leq i\leq N$, the cost function of the Agent $i$ which we denote $k^i:[0,T]\times\mathbb R^N\times\R^N\longrightarrow \R_+$. We note for any $(t,a,x)\in[0,T]\times\mathcal M_N(\R)\times\R^N$, by $k(t,a,x)\in\R^N$ the vector whose $i$th coordinate is $k^i(t,a^{:,i},x)$. Our standing assumption on the vector cost function $k$ is the following.
\begin{Assumption}\label{assump:k}
For any $(t,x)\in[0,T]\times\R^N$, the map $a\longmapsto k(t,a,x)$ is $C^1$, and has each of its coordinates which are increasing and strictly convex. Moreover, we have for some constants $C>0$ and $\ell\geq2$
$$\underset{\No{a}\rightarrow +\infty}{\underline{\lim}}\frac{\No{k(t,a,x)}}{\No{a}}=+\infty,\ \No{k(t,a,x)}\leq C\left(1+\No{a}^\ell+\No{x}\right),$$
$$\No{\nabla_a k(t,a,x)}\leq C\left(1+\No{a}^{\ell-1}\right).$$
\end{Assumption}

\vspace{0.5em}
We are now in position to define the set of admissible strategies $\mathcal A$ for the system of Agents. In our framework of study, each Agent $i$ is restricted to choose a control in a given subset $A^i$ of $\R^N$. The set  $\mathcal A$ of admissible controls is then defined as the set of $\F$-adapted processes $a$, which are $\mathcal M_N(\R)$-valued, such that for any $1\leq i\leq N$, $(a^{i,:})^\top$ takes values in $A^i$, \eqref{eq:dd} holds and 
$\int_0^Tb(s,a_s,X_s)ds$ as well as $\int_0^Tk(s,a_s,X_s)ds$ are valued in $M^\phi(\R^N)$. Recall that $M^\phi(\R^N)$ is the Morse-Transue space associated to $\R^N$, see \eqref{eq:morse}.

\subsection{The objective function of each Agent}

\vspace{0.5em}
Now that the set of admissible strategies of the agents has been established, let us turn to the design of their objective function. We assume that the Agents derive utility from two sources. First, from the salary they receive from the contract,
diminished by the cost induced by their working effort. Second, the Agents derived utility from the success of their project in comparison to the other ones. In our model, the main motivation for the interaction between the Agents comes from this feature, which makes them compare to each other.

\vspace{0.5em}
More precisely, we assume that the utilities of the Agents are exponential and that, for any $1\leq i\leq N$, given $N$ contracts $\xi:=(\xi^1,\dots,\xi^N)^\top$ and a choice of actions $a\in\mathcal A$ made by all the Agents, the utility of Agent $i$ is
\begin{equation}\label{eq:def_utility_Agents}
U_0^i(a^{:,i},a^{:,-i},\xi^i):=\E^{\P^a}\left[\mathcal U^A_i\left(\xi^i+\Gamma_i(X_T)-\int_0^Tk^i(s,a^{:,i}_s, X_s)ds\right)\right],
\end{equation}
with 
$$\mathcal U^A_i(x):=-\exp\left(-R_A^ix\right), \ x\in\R,\ \Gamma_i:\mathbb R^{N}\longrightarrow\R,$$
where $R^A_i>0$ represents the risk-aversion of Agent $i$, and the map $\Gamma_i$ corresponds to Agent $i$ comparing his performance with the performances of the other Agents. This map $\Gamma^i$ can be quite general and a typical example would be 
\begin{equation}\label{eq:examples}
\Gamma_i(x):=\gamma_i\left(x^i-\bar x^{-i}\right),\ x\in\R^N,
\end{equation}
where $\gamma_i$ is a given non-negative constant, so called competition index of Agent $i$. This setting corresponds to the case where each Agent compares his performance to the average of the other Agents performances. The higher $\gamma^i$ is, the more competitive Agent $i$ will be.

\vspace{0.5em}
In general, we assume that the comparison map $\Gamma$ satisfies 
\begin{Assumption}\label{assump:gamma}
For any $1\leq i\leq N$, the maps $\Gamma_i=\R^N\longrightarrow \R$ are $($Borel$)$ measurable and satisfy, for some $C>0$
$$\abs{\Gamma_i(x)}\leq C\left(1+\No{x}\right),\; x\in\R^N.$$
\end{Assumption}

 \subsection{The contracting setting for the Principal}
 Now that the incentives of the Agents are well understood, we can turn to the design of those of the Principal. In our setting, the Principal offers simultaneously a contract to each Agent at time $0$, and he can commit to any such contract. For any $1\leq i\leq N$, a contract for Agent $i$ will then be represented as a real valued random variable $\xi^i$, which, for now, we only assume to be $\mathcal F_T$-measurable, representing the amount of money that Agent $i$ will receive at time $T$ at the end of the contract. Observe that there are no intertemporal payments.
 
 \vspace{0.5em}
 Given such vector of terminal payment $\xi$, the system of Agents in interaction will choose some response actions $a\in\mathcal A$. Hence, each agent will obtain from the game at time $0$ the utility value $U_0^i(a,\xi^i)$. In order to ensure that each Agent agrees to enter the game, the Principal will restrict his contracts offers to those such that each agent $i$ receives at least his reservation utility denoted $\underbar{U}^i$, i.e such that
 \begin{equation}\label{reserv}
 U_0^i(a,\xi^i) \;\ge\;  \underbar{U}^i,\ 1\le i \le N.
 \end{equation}
 
Besides, for given contracts $(\xi^i)_{1\leq i\leq N}$ and a given choice of actions $a\in\mathcal A$ made by the Agents, the Principal derives utility from the terminal values of the projects as follows
\begin{equation}\label{eq:def_utility_Principal}
U_0^P(a,\xi):=\E^{\P^a}\left[-\exp\left(-R_P\left(X_T-\xi\right)\cdot {\bf 1}_N\right)\right],
\end{equation}
where $R_P>0$ is the risk-aversion of the Principal.

\vspace{0.5em}
Of course, for all of this to be meaningful, we still need to impose conditions on $a$ and $\xi$ so that \reff{eq:def_utility_Agents} and \reff{eq:def_utility_Principal} are well defined. We will take care of this problem later on when defining the set of admissible contracts.

\vspace{0.5em}
 To sum up, the Principal will choose $N$ contracts $(\xi^i)_{1\leq i\leq N}$, leading to a response effort $a$ of the system of Agents, such that each Agent wishes to enter the contract, i.e. \eqref{reserv} is satisfied, and  which should be optimal according to the enhanced criterion \eqref{eq:def_utility_Principal}. Depending on how much information is available for the Principal, we now split our study into the so-called corresponding first-best and second-best settings.

\section{The first-best problem}\label{sec:first}
In this section, we concentrate our attention to the so-called first-best framework, where there is no moral hazard and the Principal can actually choose directly both the contracts $(\xi^i)_{1\leq i\leq N}$ as well as the actions of the Agents. As far as we know, this problem in our framework has never been solved in the literature.

\vspace{0.5em}
We first rewrite the problem of the Principal in a more tractable stochastic control form. Then, we provide in Section \ref{subsec: first best 2} a representation of the optimal contract in full generality. We finally focus more closely in Section \ref{sec:gammalinear} on the more tractable setting with linear comparison map $\Gamma$ of the form  \eqref{eq:examples}. 

\subsection{Stochastic control reformulation of the Principal problem}\label{subsec: first best 1}

In this case, the set of admissible contracts $\mathcal C^{FB}$ will be defined as
$$\mathcal C^{FB}:=\left\{\xi,\text{ $\Fc_T$-measurable with $\xi\in M^\phi(\R^N)$}\right\}.$$
It is then clear by construction  that for any $(a,\xi)\in\mathcal A\times\mathcal C^{FB}$, the quantities $U_0^i(a^{:,i},a^{:,-i},\xi^i)$ and $U_0^P(a,\xi)$ are well-defined.

\vspace{0.5em}
Given some reservation utility levels $(\underbar{U}^i)_{1\leq i\leq N}$ (which are all negative) chosen by the Agents, the problem of the Principal is then
\begin{align}\label{eq:firstbest}
 U_0^{P,FB}:=\underset{a\in\mathcal A}{\sup}\ \underset{\xi\in\mathcal C^{FB}}{\sup}\left\{U_0^P(a,\xi)+\sum_{i=1}^N\rho_iU_0^i(a^{:,i},a^{:,-i},\xi^i)\right\},
\end{align}
where the $\rho_i>0$ are the Lagrange multiplier associated to the participation constraints.

\vspace{0.5em}
Let us start with the maximization with respect to $\xi$. In order to do so, for any $a\in\mathcal A$, let us consider the following map $\Xi^a:M^\phi(\R^N)\longrightarrow \mathbb R$ defined by
$$\Xi^a(\xi):=\E^{\P^a}\left[-e^{-R_P(X_T-\xi)\cdot{\bf 1}_N}-\Sum_{i=1}^N\rho_ie^{-R_A^i\left(\xi^i+\Gamma_i(X_T)-\int_0^Tk^i(s,a^{:,i}_s, X_s)ds\right)}\right].$$
Since $\rho_i>0$ for any $1\leq i\leq N$, it can easily be seen that $\Xi^a$ is continuous, strictly convex, proper, and G\^ateaux differentiable, with G\^ateaux derivative given, for any $h\in M^\phi(\R^N)$, by
\begin{align*}
D\Xi^a(\xi)[h]=&\ \E^{\P^a}\left[-R_Ph\cdot{\bf 1}_Ne^{-R_P(X_T-\xi)\cdot{\bf 1}_N}\right.\\
&\left.\hspace{2.5em}+\Sum_{i=1}^N\rho_iR_A^i h^ie^{-R_A^i\left(\xi^i+\Gamma_i(X_T)-\int_0^Tk^i(s,a^{:,i}_s,X_s)ds\right)}\right].
\end{align*}
For any $a\in\mathcal A$, let introduce $\xi^\star (a)$ by
$$(\xi^\star (a))^i:=\frac{1}{R_A^i}\log\left(\frac{\rho_iR_A^i}{R_P}\right)-\Gamma_i(X_T)+\int_0^Tk^i(s,a_s^{:,i},X_s)ds+\frac{R_P}{R_A^i}(X_T\cdot{\bf 1}_N-\xi^\star (a)\cdot{\bf 1}_N),$$
together with
\begin{align}\label{eq:focxi2}
\nonumber\xi^\star (a)\cdot{\bf 1}_N:=&\ \frac{\overline R_A}{\overline R_A+NR_P}\left(\frac{R_PN}{\overline R_A} X_T-\Gamma(X_T)+ \int_0^Tk(s,a_s,X_s)ds\right)\cdot {\bf 1}_N\\
&+\frac{\overline R_A}{\overline R_A+NR_P}\sum_{i=1}^N\frac{1}{R_A^i}\log\left(\frac{\rho_iR_A^i}{R_P}\right),
\end{align}
where $\Gamma(X_T)\in\R^N$ is the vector whose $i$th coordinate is $\Gamma_i(X_T)$, and where 
$$\overline R_A:=\frac{N}{\sum_{i=1}^N\frac{1}{R_A^i}}.$$
Then, for any $h\in M^\phi(\R^N)$, we have $D\Xi^a(\xi^\star (a))[h]=0$, so that this $\xi^\star (a)$ attains the minimum of $\Xi^a$ and is therefore optimal.

\vspace{0.5em}
Plugging these expressions back into the Principal problem, we obtain that the principal problem rewrites in a stochastic control form as 
\begin{align*}
\frac{\overline R_A+NR_P}{\overline R_A}\Prod_{i=1}^N\left[\left(\frac{\rho_iR_A^i}{R_P}\right)^{\frac{R_P\overline R_A}{R_A^i(\overline R_A+NR_P)}}\right]\underset{a\in\mathcal A}{\sup}\ \E^{\P^a}\left[-e^{\frac{R_P\overline R_A}{\overline R_A+NR_P}\left( \int_0^Tk(s,a_s,X_s)ds-X_T-\Gamma(X_T)\right)\cdot {\bf 1}_N}\right].
\end{align*}

We will first consider the case of a general interaction function $\Gamma$ and then tackle a benchmark case where $\Gamma$ is linear, for which the solution is much easier to find.

\subsection{General comparison map  $\Gamma$}\label{subsec: first best 2}


\subsubsection{A BSDE solution for the Principal problem}
Under the form above, the problem of the Principal is now a classical stochastic control problem (under weak formulation), with controlled state process $X$, whose drift only is controlled. Denote by $A:=\prod_{i=1}^NA^i$. It is then a classical result (see for instance \cite{elkarq,ekpq}, or similar comparison arguments in Section \ref{sec:bestreac} below) that we have
$$U_0^{P,FB}=-\frac{\overline R_A+NR_P}{\overline R_A}\Prod_{i=1}^N\left[\left(\frac{\rho_iR_A^i}{R_P}\right)^{\frac{R_P\overline R_A}{R_A^i(\overline R_A+NR_P)}}\right]\exp\left(-\frac{R_P\overline R_A}{\overline R_A+NR_P}Y_0\right),$$
where $(Y,Z)$ denotes the maximal solution of the following BSDE
\begin{align}\label{eq:bsdefb}
Y_t=\left(X_T+\Gamma(X_T)\right)\cdot {\bf 1}_N+\int_t^TF(s,X_s,Z_s)ds-\int_t^TZ_s\cdot \Sigma_sdW_s,\ t\in[0,T],\ \P-a.s.,
\end{align}
where the generator $F$ is given by
$$F(t,x,z):=\underset{a\in A}{\sup}\left\{b(t,a,x)\cdot z-k(t,a,x)\cdot {\bf 1}_N\right\}-\frac{R_P\overline R_A}{2(\overline R_A+NR_P)}\No{\Sigma_tz}^2.$$
We still need to justify why the BSDE \eqref{eq:bsdefb} indeed admits a maximal solution. First, by Assumptions \ref{assump:sigma} and \ref{assump:gamma}, we know that the terminal condition has linear growth w.r.t. $X_T$ and thus admits exponential moments of any order under $\P$ (remember that, under $\P$, $X_T$ is Gaussian). Moreover, by Lemma \ref{lemma:fgrowth} below, we have that
$$\abs{\underset{a\in A}{\sup}\left\{b(t,a,x)\cdot z-k(t,a,x)\cdot {\bf 1}_N\right\}}\leq C\left(1+\No{z}^{\frac{\ell}{\ell -1}}\right).$$
Hence, since $\ell\geq 2$, $F$ has at most quadratic growth in $z$. Since $F(t,x,0)=0$, the existence of a solution is given by \cite{bh}. However, sufficient (but not necessary\footnote{Another condition would be that $F$ is actually purely quadratic in $z$, that is to say
$$F(t,x,z)=f(t,x)+\No{\gamma(t)z}^2,$$
for some maps $f:[0,T]\times \R^N\longrightarrow \R$ and $\gamma:[0,T]\longrightarrow \Mc_N(\R)$, in which case one can easily show, using an exponential transformation, that the BSDE actually has a unique solution. This would be the case if for instance $b$ was chosen linear in $a$ and $k$ appropriately quadratic in $a$ (see the next section).}) condition for the existence of a maximal solution is that $F$ is bounded from above by a map which has linear growth in $z$ (see \cite{bek}), or that $F$ is concave or convex in $z$ (see \cite{bh2}). Such a condition requires the following additional assumption
\begin{Assumption}\label{assump:max}
Either the map $z\longmapsto F(t,x,z)$ is convex or concave for any $(t,x)\in[0,T]\times\mathbb R^N$, or there exists some $C>0$ such that for any $(t,x,z)\in[0,T]\times\R^N\times\R^N$
$$F(t,x,z)\leq C\left(1+\No{x}+\No{z}\right).$$
\end{Assumption}
\begin{Remark}
Whenever the set $A$ is compact, then Assumption \ref{assump:max} is automatically satisfied.
\end{Remark}

Denote, for any $(t,z,x)\in[0,T]\times\R^N\times\R^N$, by $\tilde a^\star (t,x,z)\in\mathcal M_N$ one of the maximizers of the map $a\longmapsto b(t,a,x)\cdot z-{\bf 1}_N\cdot k(t,a,x)$. Notice that such a maximizer is well defined since by Assumptions \ref{assump:b} and \ref{assump:k}, $k$ has superlinear growth at infinity, while $b$ has linear growth (see \reff{eq:growthb}), so that the map considered here is coercive. By a classical measurable selection argument, we deduce that the corresponding predictable process $a^\star _t:=\tilde a^\star (t,X_t,Z_t)$, defined $dt\times d\P-a.e.$, is the optimal effort for the Agents, chosen by the Principal, and that the corresponding contract $\xi^\star (a^\star )$ is optimal, provided they are both admissible. So as not to complicate further our presentation, we refrain from giving general conditions under which this holds true.

\vspace{0.5em}
Using this contract and this action, one still needs to choose the Lagrange multipliers $\rho_i$ such that
\begin{align}\label{eq:lag}
&U_0^i((a^\star )^{:,i},(a^\star )^{:,-i},(\xi^\star (a^\star ))^i)=\underbar{U}^i.
\end{align}

We have thus obtained the following verification type result
\begin{Theorem}
Let Assumptions \ref{assump:sigma}, \ref{assump:b}, \ref{assump:k}, \ref{assump:gamma} and \ref{assump:max} hold and assume furthermore that $a^\star \in\mathcal A$ and $\xi^\star (a^\star )\in\mathcal C^{FB}$. Then, the contract $\xi^\star (a^\star )$ chosen so that \reff{eq:lag} holds is optimal for the Principal, with a recommended level of effort $a^\star$.
\end{Theorem}

Instead of elaborating further on technical conditions ensuring the admissibility of the effort and contract, we choose to specialize to a linear-quadratic framework with simpler dynamics, for which the BSDE \eqref{eq:bsdefb} can be solved explicitly.

\subsubsection{Resolution in a linear-quadratic setting}
We now consider a simplified linear-quadratic setting, i.e. where the drift of the output process and the cost function are respectively linear and quadratic with respect to the control $a$. Namely, we work under the additional assumption: 
\begin{Assumption} \label{assump:truc}
For any $1=1,\dots,N$, we have $A^i=\R^N$, 
the maps $b^i$ and $k^i$ have the form
$$b^i(t,a,x)=B{\bf 1}_N\cdot a+\tilde b^i(t,x),\ k^i(t,a,x)=\frac{K}{2}\No{a}^2+\tilde k^i(t,x),\ \forall(t,a,x)\in[0,T]\times\R^N\times\R^N,$$
for some $B\in \R$, $K\in(0,+\infty)$, and some maps $\tilde b^i:[0,T]\times\R^N\longrightarrow \R$ and $
\tilde k^i:[0,T]\times\R^N\longrightarrow \R_+$. Moreover, we assume that the volatility matrix $\Sigma_t$ does not depend on time and is a multiple of the identity matrix, that is $\Sigma_t=\sigma I_N$, for some $\sigma\in (0,+\infty)$.
\end{Assumption}

\begin{Remark}
Observe in particular that Assumption \ref{assump:max} is a direct consequence of Assumption \ref{assump:truc}. Similarly the linear upper bound on the driver $F$ required in Assumption \ref{assump:max} is automatically valid if Assumptions \ref{assump:b},  \ref{assump:k}, and  \ref{assump:truc} are satisfied.
\end{Remark}

As usual, we denote by $\tilde b(t,x)$ (resp. $\tilde k(t,x)$) the vector of $\R^N$ whose $i$th component is $\tilde b^i(t,x)$ (resp. $\tilde k^i(t,x)$). Under Assumption \ref{assump:truc}, direct computations show that
$$F(t,x,z)=N\left(\frac{\abs{B}^2}{k}-\abs{\sigma}^2\frac{R_P\overline R_A}{\overline R_A+NR_P}\right)\sum_{i=1}^N\frac{\abs{z^i}^2}{2}+\tilde b(t,x)z-\tilde k(t,x)\;,$$     
which corresponds to an optimal effort $a^\star(z)\in\Mc_N(\R)$ such that
$$a^\star(z):=\frac{B}{k}z{\bf 1}_N^\top,\ z\in\R^N.$$
 Recall that $(Y,Z)$ is the maximal solution to the BSDE  \eqref{eq:bsdefb} and define
$$\eta:=N\left(\frac{\abs{B}^2}{k\abs{\sigma}^2}-\frac{R_P\overline R_A}{\overline R_A+NR_P}\right),\ P_t:=\exp\left(\eta Y_t\right),\ Q_t:=\sigma\eta P_tZ_t,\ t\in[0,T].$$
A simple application of It\^o's formula together with \eqref{eq:bsdefb} gives
$$P_t=e^{\eta(X_T+\Gamma(X_T))\cdot {\bf 1}_N}+\int_t^T\left(\frac{\tilde b(s,X_s)}{\sigma} \cdot Q_s-\eta P_s\tilde k(s, X_s)\cdot {\bf 1}_N\right)ds-\int_t^TQ_s\cdot dW_s,$$
that is $(P,Q)$ solves a simple linear BSDE. In particular, defining a new probability measure $\widetilde \P$ by
$$\frac{d\widetilde \P}{d\P}:=\Ec\left(\int_0^T\frac{\tilde b(s,X_s)}{\sigma}\cdot dW_s\right),$$
we rewrite
$$P_t=\E^{\widetilde \P}\left[\left.\exp\left(\eta{\bf 1}_N\cdot\left(X_T+\Gamma(X_T)-\int_t^T\tilde k(s,X_s)ds\right)\right)\right|\Fc_t\right],$$
so that
$$Y_t=\frac{1}{\eta}\log\left(\E^{\widetilde \P}\left[\left.\exp\left(\eta{\bf 1}_N\cdot\left(X_T+\Gamma(X_T)-\int_t^T\tilde k(s,X_s)ds\right)\right)\right|\Fc_t\right]\right).$$
In order to have access to the optimal effort $a^\star$, we need to reinforce Assumption \ref{assump:gamma}.
\begin{Assumption}\label{assump:gamma2}
The Borel measurable map $\Gamma$ is Lipschitz-continuous.
\end{Assumption}
Then, it is a well known result (see for instance Proposition 5.3 in \cite{ekpq}), that a version of $Q_t$ is given by $D_tP_t$, where $D$ is the Malliavin differentiation operator. Since $P$ is given as a conditional expectation of the composition of a smooth and a Lipschitz-continuous function, we compute directly using the chain rule of Malliavin calculus that
\begin{align*}
Q_t=&\ \E^{\widetilde \P}\left[\eta\left(D_tX_T+\Gamma^{'}(X_T)\cdot{\bf 1}_N D_tX_T-\int_t^T\tilde k_x(s,X_s)\cdot {\bf 1}_ND_tX_sds\right)\right.\\
&\left.\left.\hspace{1.8em}\times \exp\left(\eta{\bf 1}_N\cdot\left(X_T+\Gamma(X_T)-\int_t^T\tilde k(s,X_s)ds\right)\right)\right|\Fc_t\right]\\
&+\E^{\widetilde \P}\left[\left.\left(\frac1\sigma\int_t^T\tilde b_x(s,X_s)D_tX_sdW_s-\frac1{\sigma^2}\int_t^T\tilde b_x(s,X_s)\tilde b(s,X_s)D_tX_s\right)\right.\right.\\
&\left.\left.\hspace{2.7em}\times \exp\left(\eta{\bf 1}_N\cdot\left(X_T+\Gamma(X_T)-\int_t^T\tilde k(s,X_s)ds\right)\right)\right|\Fc_t\right],
\end{align*}
where for any $t\in[0,T]$ and Lebesgue almost every $x\in\R^N$, $\Gamma^{'}(x)$ denotes the vector of $\R^N$ whose $i$th component is $(\Gamma^i)^{'}(x)$ and $\tilde k_x(t,x)$ denotes the vector of $\R^N$ whose $i$th component is $\tilde k_x^i(t,x)$. We deduce that
\begin{align*}
a^\star(Z_t)=&\ \frac{B}{k\eta}\left(\E^{\widetilde \P}\left[\eta\left(D_tX_T+\Gamma^{'}(X_T)\cdot{\bf 1}_N D_tX_T-\int_t^T\tilde k_x(s,X_s)\cdot {\bf 1}_ND_tX_sds\right)\right.\right.\\
&\left.\left.\left.\hspace{3.9em}\times \exp\left(\eta{\bf 1}_N\cdot\left(X_T+\Gamma(X_T)-\int_t^T\tilde k(s,X_s)ds\right)\right)\right|\Fc_t\right]\right.\\
&\left.\hspace{1.8em}+\E^{\widetilde \P}\left[\left.\left(\frac1\sigma\int_t^T\tilde b_x(s,X_s)D_tX_sdW_s-\frac1{\sigma^2}\int_t^T\tilde b_x(s,X_s)\tilde b(s,X_s)D_tX_s\right)\right.\right.\right.\\
&\left.\left.\left.\hspace{3.9em}\times \exp\left(\eta{\bf 1}_N\cdot\left(X_T+\Gamma(X_T)-\int_t^T\tilde k(s,X_s)ds\right)\right)\right|\Fc_t\right]\right)\\
&\times\left(\E^{\widetilde \P}\left[\left.\exp\left(\eta{\bf 1}_N\cdot\left(X_T+\Gamma(X_T)-\int_t^T\tilde k(s,X_s)ds\right)\right)\right|\Fc_t\right]\right)^{-1}{\bf 1}_N^\top.
\end{align*}
Finally, if $\tilde k$ and $\tilde b$ do not depend on $x$, we can further simplify the above expression, using the fact that under $\widetilde \P$, $X_T$ is, conditionally on $\Fc_t$, an $N$-dimensional Gaussian random variable with mean $X_t+\int_t^T\tilde b(s)ds$ and variance-covariance matrix $\sigma^2(T-t) I_N$
\begin{align*}
Y_t=&\ X_t \cdot {\bf 1}_N - \int_t^T\tilde k(s)\cdot {\bf 1}_Nds-\frac{N}{2\eta}\log\left(2\pi\sigma^2(T-t)\right)\\
&+\frac1\eta\log\left(\int_{\R^N}e^{\eta {\bf 1}_N\cdot\left(x+\Gamma(x+X_t)\right)-\frac{\No{x-\int_t^T\tilde b(s)ds}^2}{2\sigma^2(T-t)}}dx\right).
\end{align*}
Similarly, we have
\begin{align*}
Q_t=&\ \eta\sigma{\bf 1}_N\int_{\R^N}\left(1+\Gamma^{'}(x+X_t)\cdot{\bf 1}_N\right) e^{\eta{\bf 1}_N\cdot\left(x+X_t+\Gamma(x+X_t)-\int_t^T\tilde k(s)ds-\frac{\No{x-\int_t^T\tilde b(s)ds}^2}{2\sigma^2(T-t)}\right)}dx,
\end{align*}
so that
\begin{align*}
a^\star(Z_t)=&\ \frac{B\sigma}{k}\frac{\displaystyle\int_{\R^N}\left(1+\Gamma^{'}(x+X_t)\cdot{\bf 1}_N\right) e^{\eta{\bf 1}_N\cdot\left(x+X_t+\Gamma(x+X_t)-\int_t^T\tilde k(s)ds-\frac{\No{x-\int_t^T\tilde b(s)ds}^2}{2\sigma^2(T-t)}\right)}dx}{\displaystyle\int_{\R^N} e^{\eta{\bf 1}_N\cdot\left(x+X_t+\Gamma(x+X_t)-\int_t^T\tilde k(s)ds-\frac{\No{x-\int_t^T\tilde b(s)ds}^2}{2\sigma^2(T-t)}\right)}dx}.
\end{align*}
We summarize all the above in the following theorem.

\begin{Theorem}
Let Assumptions 
\ref{assump:b}, \ref{assump:k}, 
\ref{assump:truc} and \ref{assump:gamma2} hold. Then, if the contract $\xi^\star(a^*)$ $($where the $\rho^i$ are chosen so that \eqref{eq:lag} holds$)$ belongs to $\Cc^{FB}$, this contract is optimal for the Principal, where the optimal effort $a^\star$ is given by the following process
\begin{align*}
a^\star_t:=&\ \frac{B}{k\eta}\left(\E^{\widetilde \P}\left[\eta\left(D_tX_T+\Gamma^{'}(X_T)\cdot{\bf 1}_N D_tX_T-\int_t^T\tilde k_x(s,X_s)\cdot {\bf 1}_ND_tX_sds\right)\right.\right.\\
&\left.\left.\left.\hspace{3.9em}\times \exp\left(\eta{\bf 1}_N\cdot\left(X_T+\Gamma(X_T)-\int_t^T\tilde k(s,X_s)ds\right)\right)\right|\Fc_t\right]\right.\\
&\left.\hspace{1.8em}+\E^{\widetilde \P}\left[\left.\left(\frac1\sigma\int_t^T\tilde b_x(s,X_s)D_tX_sdW_s-\frac1{\sigma^2}\int_t^T\tilde b_x(s,X_s)\tilde b(s,X_s)D_tX_s\right)\right.\right.\right.\\
&\left.\left.\left.\hspace{3.9em}\times \exp\left(\eta{\bf 1}_N\cdot\left(X_T+\Gamma(X_T)-\int_t^T\tilde k(s,X_s)ds\right)\right)\right|\Fc_t\right]\right)\\
&\times\left(\E^{\widetilde \P}\left[\left.\exp\left(\eta{\bf 1}_N\cdot\left(X_T+\Gamma(X_T)-\int_t^T\tilde k(s,X_s)ds\right)\right)\right|\Fc_t\right]\right)^{-1}{\bf 1}_N^\top.
\end{align*}
which is simplified, when $\tilde k$ and $\tilde b$ do not depend on $x$ to
\begin{align*}
a^\star_t=&\ \frac{B\sigma}{k}\frac{\displaystyle\int_{\R^N}\left(1+\Gamma^{'}(x+X_t)\cdot{\bf 1}_N\right) e^{\eta{\bf 1}_N\cdot\left(x+X_t+\Gamma(x+X_t)-\int_t^T\tilde k(s)ds-\frac{\No{x-\int_t^T\tilde b(s)ds}^2}{2\sigma^2(T-t)}\right)}dx}{\displaystyle\int_{\R^N} e^{\eta{\bf 1}_N\cdot\left(x+X_t+\Gamma(x+X_t)-\int_t^T\tilde k(s)ds-\frac{\No{x-\int_t^T\tilde b(s)ds}^2}{2\sigma^2(T-t)}\right)}dx}.
\end{align*}
\end{Theorem}

It is worth noticing that our framework allow for the consideration of rather general $\Gamma$ functions, allowing for example to consider Agents interested in their (smoothed approximate) ranking within the population of Agents. Nevertheless, in order to obtain more explicit representation of the solution, we now focus on a particular form of comparison criterium in between the Agents, relying on a linear $\Gamma$ function.

\subsection{Average relative benchmark and linear comparison map $\Gamma$}\label{sec:gammalinear}
In this section, we do not focus anymore on a linear-quadratic setting but specialize the discussion to another solvable framework by imposing:
\begin{Assumption}\label{assump:example}
We have
$$\Gamma_i(x)=x^i-\bar x^{-i},\; b^i(t,a,x)= b^i(t,a),\; k^i(t,a,x)= k^i(t,a),\; 1\leq i\leq N,\; (a,x)\in\R^N\times\R^N.$$
\end{Assumption}

In this case, we have
$$\Gamma(X_T)\cdot {\bf 1}_N=\left(\gamma-\overline \gamma ^-\right)\cdot X_T,$$
where $\gamma\in\R^N$ is the vector whose $i$th coordinate is $\gamma_i$ and where $\overline \gamma^-$ is such that  for any $1\leq i\leq N$, $(\overline \gamma ^-)^i:=\overline \gamma^{-i}$.

\subsubsection{Explicit solution for the Principal problem}

\vspace{0.5em}
Denote $p:=\overline\gamma^--{\bf 1}_N-\gamma$. The Principal problem can then be rewritten
\begin{align*}
&\frac{\overline R_A+NR_P}{\overline R_A}e^{\frac 12\left(\frac{R_P\overline R_A}{\overline R_A+NR_P}\right)^2\int_0^T \No{\Sigma_t p}^2dt} \Prod_{i=1}^N\left[\left(\frac{\rho^iR_A^i}{R_P}\right)^{\frac{R_P\overline R_A}{R_A^i(\overline R_A+NR_P)}}\right]\\
&\hspace{0.9em}\times\underset{a\in\mathcal A}{\sup}\ \E^{\P^a}\left[-\mathcal E\left(\frac{R_P\overline R_A}{\overline R_A+NR_P}p\cdot \int_0^T \Sigma_t dW^a_t\right)e^{\frac{R_P\overline R_A}{\overline R_A+NR_P}\int_0^T\left(p\cdot b(s,a_s)+{\bf 1}_N\cdot k(s,a_s)\right)ds}\right].
\end{align*}
Let $a^\star (t)$ be any (deterministic) minimizer of the map $a\longmapsto p\cdot b(t,a)+{\bf 1}_N\cdot k(t,a)$. Since $W^a$ is a Brownian motion under $\P^a$, it is clear that the stochastic exponential which appears above is a uniformly integrable martingale. Hence, we deduce that
\begin{align*}
U_0^{P,FB}\leq&\ \frac{\overline R_A+NR_P}{\overline R_A}e^{\frac 12\left(\frac{R_P\overline R_A}{\overline R_A+NR_P}\right)^2\int_0^T\No{\Sigma_t p}^2dt}\Prod_{i=1}^N\left[\left(\frac{\rho^iR_A^i}{R_P}\right)^{\frac{R_P\overline R_A}{R_A^i(\overline R_A+NR_P)}}\right]\\
&\times \exp\left(\frac{R_P\overline R_A}{\overline R_A+NR_P}\int_0^T\left(p\cdot b(s,a^\star_s)+{\bf 1}_N\cdot k(s,a^\star_s)\right)ds\right).
\end{align*}
But this upper bound can easily be attained by choosing action $a^\star $ and contract $\xi^\star (a^\star )$ as in \eqref{eq:focxi2}. Moreover, since $a^\star $ is deterministic, it obviously belongs to $\mathcal A$, and we also have $\xi^\star (a^\star )\in\mathcal C^{FB}$, since it is linear in $X$ which has exponential moments of any order.

\vspace{0.5em} 
Finally, let us compute the utility that Agent $i$ obtains from this contract. Recalling \eqref{eq:def_utility_Agents}, we get 
\begin{align*}
&U_0^i((a^\star )^{:,i},(a^\star )^{:,-i},(\xi^\star (a^\star ))^i)\\
=&\ \frac{R_P}{\rho_iR_A^i}\Prod_{j=1}^N\left[\left(\frac{\rho_jR_A^j}{R_p}\right)^{\frac{R_P\overline R_A}{R_A^j(\overline R_A+NR_P)}}\right]e^{\frac{R_P\overline R_A}{\overline R_A+NR_P}{\bf 1}_N\cdot\int_0^Tk(s,a^\star_s)ds}\\
&\times\E^{\P^{a^\star }}\left[-\exp\left(\frac{R_P\overline R_A}{\overline R_A+NR_P}p\cdot X_T\right)\right]\\
=&\ -\frac{R_P}{\rho_iR_A^i}\Prod_{j=1}^N\left[\left(\frac{\rho_jR_A^j}{R_p}\right)^{\frac{R_P\overline R_A}{R_A^j(\overline R_A+NR_P)}}\right]e^{\frac{R_P\overline R_A}{\overline R_A+NR_P}{\bf 1}_N\cdot\int_0^Tk(s,a^\star_s)ds+\frac{R_P\overline R_A}{\overline R_A+NR_P}\int_0^Tp\cdot b(s,a^\star_s)ds}\\
&\times e^{\frac 12\left(\frac{R_P\overline R_A}{\overline R_A+NR_P}\right)^2\int_0^T\No{\Sigma_t p}^2dt}.
\end{align*}
We therefore need to determine the Lagrange multipliers $(\rho_i)_{1\leq i\leq N}$ so that we have
\begin{equation}\label{eq:lagmult}
\Prod_{1\leq j\leq N,\ j\neq i}\left[\rho_j^{\frac{R_P\overline R_A}{R_A^j(\overline R_A+NR_P)}}\right]\rho_i^{\frac{R_P\overline R_A}{R_A^i(\overline R_A+NR_P)}-1}=-\frac{R_A^i}{R_P}A\underbar{U}^i,\ \qquad  1\leq i\leq N,
\end{equation}
where $A>0$ is defined by
$$A:=e^{-\frac{R_P\overline R_A}{\overline R_A+NR_P}{\bf 1}_N\cdot\int_0^Tk(s,a^\star_s)ds-\frac{R_P\overline R_A}{\overline R_A+NR_P}\int_0^Tp\cdot b(s,a^\star_s)ds-\frac 12\left(\frac{R_P\overline R_A}{\overline R_A+NR_P}\right)^2\int_0^T\No{\Sigma_t p}^2dt}.$$
Then, if we define vectors $(B,\overline R,\log(\rho))\in\R^N\times \R^N\times\R^N$ with
$$B^i:=\frac{\overline R_A+NR_P}{R_P\overline R_A}\log\left(-\frac{R_A^i}{R_P}A\underbar{U}^i\right),\ \overline R^i:=\frac{1}{R_A^i},\ \log(\rho)^i:=\log(\rho^i),$$
then by taking logarithms on both sides of \reff{eq:lagmult}, we obtain that \reff{eq:lagmult} is actually equivalent to solving the linear system
$$\left({\bf 1}_N\overline R^\top-\frac{\overline R_A+NR_P}{R_P\overline R_A}I_N\right)\log(\rho)=B.$$
It can then be checked directly that
$$\left({\bf 1}_N\overline R^\top-\frac{\overline R_A+NR_P}{R_P\overline R_A}I_N\right)^{-1}=-\frac{R_P\overline R_A}{\overline R_A+NR_P}\left(R_P{\bf 1}_N\overline R^\top+I_N\right).$$
Therefore, we finally have 
$$\rho_i=-\frac{R_P}{AR_A^iU_i}\Prod_{j=1}^N\left[-\frac{R_P}{AR_A^j\underbar{U}^j}\right]^{\frac{R_P}{R_A^j}},\ 1\leq i\leq N.$$
We have just proved the following

\begin{Theorem}
Let Assumptions \ref{assump:sigma}, \ref{assump:b}, \ref{assump:k} and \ref{assump:example} hold. Then an optimal contract $\xi_{FB}\in\mathcal C^{FB}$ in the problem \reff{eq:firstbest}, with reservation utilities $(\underbar{U}^i)_{1\leq i\leq N}\in(-\infty,0)^N$, is given, for $i=1,\dots,N$, by
\begin{align}\label{eq:optcontractFB}
\nonumber\xi^i_{FB}:=&\ \frac{R_P\overline R_A}{R_A^i(\overline R_A+NR_P)}({\bf 1}_N+\gamma-\overline\gamma^-)\cdot X_T-\gamma_i\left(X^i_T-\overline X^{-i}_T\right)+\int_0^Tk^i(s,(a^\star )_s^{:,i})ds\\
\nonumber &+\frac{R_P\overline R_A}{R_A^i(\overline R_A+NR_P)}\int_0^Tb(s,a^\star _s)\cdot (\overline\gamma^--\gamma-{\bf 1}_N)ds-\frac{1}{R_A^i}\log(-\underbar{U}^i)\\
&+\frac{1}{2R_A^i}\left(\frac{R_P\overline R_A}{\overline R_A+NR_P}\right)^2\int_0^T\No{\Sigma_s (\overline\gamma^--\gamma-{\bf 1}_N)}^2ds.
\end{align}
where for any $t\in[0,T]$, the optimal action $a^\star _t\in\mathcal M_N(\R)$ is any minimizer of the map $a\longmapsto (\overline\gamma^--\gamma-{\bf 1}_N)\cdot b(t,a)+{\bf 1}_N\cdot k(t,a)$. Moreover, the value function of the Principal is then
$$U_0^{P,FB}=-\Prod_{i=1}^N\left[\left(-\underbar{U}^i\right)^{-\frac{R_P}{R_A^i}}\right]e^{\frac{R_P^2\overline R_A}{2(\overline R_A+NR_P)}\int_0^T\No{\Sigma_s (\overline\gamma^--\gamma-{\bf 1}_N)}^2ds+R_P\int_0^T\left(k\cdot {\bf 1}_N-b\cdot({\bf 1}_N+\gamma-\overline\gamma^-)(s,a^\star _s)\right)ds}.$$
\end{Theorem}
\subsubsection{A bidimensional linear quadratic benchmark case}\label{sec:bench}
We specialize here the discussion to a setting where everything can be computed explicitly, in particular the optimal actions of the Agents. For simplicity, we choose $N=2$, $A_1=A_2=\R^2$, as well as
$$b(t,a):=\begin{pmatrix}
a^{1,1}-a^{1,2}\\
a^{2,2}-a^{2,1}
\end{pmatrix},\ \text{for any $a:=\begin{pmatrix}a^{1,1} & a^{1,2}\\
a^{2,1} & a^{2,2}\end{pmatrix}\in\mathcal M_2(\R),$}$$
and for some constants $(k^{1,1},k^{1,2},k^{2,1},k^{2,2})\in(\R_+^\star )^4$
$$k(t,a):=\begin{pmatrix}
\frac{k^{1,1}}{2}\abs{a^{1,1}}^2+\frac{k^{2,1}}{2}\abs{a^{2,1}}^2\\
\frac{k^{2,2}}{2}\abs{a^{2,2}}^2+\frac{k^{1,2}}{2}\abs{a^{1,2}}^2
\end{pmatrix},\ \text{for any $a:=\begin{pmatrix}a^{1,1} & a^{1,2}\\
a^{2,1} & a^{2,2}\end{pmatrix}\in\mathcal M_2(\R).$}$$
In this setting, the vector $p$ is simply given by
$$p=\begin{pmatrix}
\gamma_2-\gamma_1-1\\
\gamma_1-\gamma_2-1
\end{pmatrix}.$$

\subsubsection*{Optimal effort of the agents}
In our context, the strictly convex map that we need to minimize is 
\begin{align*}
f(a):=&\ \frac{k^{1,1}}{2}\abs{a^{1,1}}^2+\frac{k^{2,1}}{2}\abs{a^{2,1}}^2+\frac{k^{2,2}}{2}\abs{a^{2,2}}^2+\frac{k^{1,2}}{2}\abs{a^{1,2}}^2-(1+\gamma_1-\gamma_2)(a^{1,1}-a^{2,2})\\
&-(1+\gamma_2-\gamma_1)(a^{1,2}-a^{2,1}).
\end{align*}
We have
\begin{align*}
\frac{\partial f}{\partial a^{1,1}}=k^{1,1}a^{1,1}-1-\gamma_1+\gamma_2,\ \frac{\partial f}{\partial a^{2,1}}=k^{2,1}a^{2,1}+1+\gamma_2-\gamma_1,\\
\frac{\partial f}{\partial a^{2,2}}=k^{2,2}a^{2,2}+1+\gamma_1-\gamma_2,\ \frac{\partial f}{\partial a^{1,2}}=k^{1,2}a^{1,2}-1-\gamma_2+\gamma_1,
\end{align*}
so that the optimal actions of the two Agents are
$$a^\star :=\begin{pmatrix}
\frac{1+\gamma_1-\gamma_2}{k^{1,1}} &\frac{1+\gamma_2-\gamma_1}{k^{1,2}}\\
-\frac{1+\gamma_2-\gamma_1}{k^{2,1}} & -\frac{1+\gamma_1-\gamma_2}{k^{2,2}}
\end{pmatrix}.$$

Hence, if for instance Agent $1$ is much more competitive than Agent $2$, so that $\gamma_1>1+\gamma_2$, then Agent $1$ will work towards his project and will also work to decrease the value of the project of Agent $2$, while Agent $2$ will work to decrease the value of his own project and to increase the value of the project of Agent $1$.

\subsubsection*{Optimal recruitment scheme for the principal}

Let us now ask ourselves the question of the optimal type of Agents that the Principal should hire. More precisely, given the choice between many Agents, what are the optimal parameters $(\gamma_1,\gamma_2)$ for the Principal?

\vspace{0.5em}
From our general result, the problem for the Principal is then to maximize his value function, which is then equivalent to minimizing the map
$$g(\gamma_1,\gamma_2):=(1+\gamma_1-\gamma_2)^2\alpha_1+(1+\gamma_2-\gamma_1)^2\alpha_2,$$
where
$$\alpha_1:=\frac{R_P\overline R_A}{\overline R_A+2R_P}\sigma^2_1-\left(\frac{1}{k^{1,1}}+\frac{1}{k^{2,2}}\right),\ \alpha_2:=\frac{R_P\overline R_A}{\overline R_A+2R_P}\sigma^2_2-\left(\frac{1}{k^{1,2}}+\frac{1}{k^{2,1}}\right).$$ 
Then, it is clear that if $\alpha_1+\alpha_2\leq0$, the Principal would like to hire Agents with $\abs{\gamma_1-\gamma_2}\longrightarrow +\infty$, whereas if $\alpha_1+\alpha_2>0$, the Principal wants to hire Agents with
$$\gamma_1-\gamma_2=\frac{\alpha_2-\alpha_1}{\alpha_1+\alpha_2}.$$
Notice that this situation is optimal for the Principal, but also for the hired Agents, since in any case they receive their reservation utility. More importantly, let us emphasize that in our model, the principal should hire agents with different competitively appetence profile. A firm has economic gain from hiring Agents with diverse competitive profiles. 

\subsubsection{Economic interpretation}
The general from of the optimal contract that we have obtained is given by
$$(\xi^\star )^i(a^\star )=C_i+\frac{R_P\overline R_A}{R_A^i(\overline R_A+NR_P)}({\bf 1}_N+\gamma-\overline\gamma^-)\cdot X_T-\gamma_i\left(X^i_T-\overline X^{-i}_T\right),$$
for some constant $C_i$, allowing to satisfy the participation constraint.

\vspace{0.5em}
Hence, the Principal penalizes each Agent with the amount $-\gamma_i(X^i_T-\overline X^{-i}_T)$, so as to suppress the appetence for competition of the Agents. More precisely, Agents who do better than the average are penalized, while Agents who do worse are gratified, with the exact amount which makes them indifferent towards the performances of the other Agents. Competitive Agents are paid less but each other Agent has incentives to work on the project of a competitive Agent. Moreover, each Agent is paid a positive part of each projects, the percentage depending on the risk aversion of the Agent, and of the universal vector
$$\frac{R_P\overline R_A}{\overline R_A+NR_P}({\bf 1}_N+\gamma-\overline\gamma^-).$$
Hence, each Agent receives, for any $1\leq i\leq N$, a fraction of the terminal value of the $i$th project, which is proportional to $1+\gamma_i-\overline \gamma^{-i}$. This therefore means that if an Agent is particularly competitive, compared to the others, then all the Agents will receive a large part of his project, which gives them incentives not to work against the interest of this particular Agent. Conversely, if Agent $i_0$ is not very competitive, then the other Agents could be penalized (if $1+\gamma_{i_0}-(\overline\gamma^{-})^{i_0}<0$) by the terminal value of his project, which gives them incentives to reduce the value of his project as much as possible.

\vspace{0.5em}
 Observe that the objective function of a competitive Agent is such that, whenever his project succeeds (in comparison to the others), he requires less salary for a similar utility value of the game at time $0$. Hence, we observe that a Principal should allocate competitive Agents to projects with the highest probability of success, i.e. those with smallest volatility and those which may benefit from the help of the other Agents. Similarly, it is worth noticing that it is in the firm (i.e. the Principal) interest to hire Agents with diverse competition appetence profiles. 

\section{The second-best problem}\label{sec:second}
In this section, we consider the so-called second-best problem when moral hazard exists. In such a case, the Principal cannot observe the actions chosen by the Agents, and can only control the salaries that he offers. The main difficulty here compared to the one-Agent case of Holmstr\"om and Milgrom \cite{hm}, is that, given a contract $\xi$, we have to be able to find an equilibrium resulting of the interactions between the Agents. Since the agents are playing simultaneously, we are looking for a Nash equilibrium.

\subsection{Nash equilibria for the Agents}
\subsubsection{Definition and assumption}
The notion of equilibrium of interest is that of a Nash equilibrium. This calls for a first definition. For any $i=1,\dots,N$, and for any action $a^{-i}$ valued in $\mathcal M_{N,N-1}$ chosen by all the Agents $j\neq i$, we define the set
$$\mathcal A^i(a^{-i}):=\left\{(a_s)_{0\leq s\leq T},\ \text{$\R^N$-valued, such that $a\otimes_ia^{-i}\in\mathcal A$}\right\}.$$

\begin{Definition}
Given a contract $\xi$, a Nash equilibrium for the $N$ Agents is an action $a^\star (\xi)\in\mathcal A$ such that for any $i=1,\dots, N$, we have
$$\underset{a\in\mathcal A^i((a^\star )^{:-i}(\xi))}{\sup} U^i_0(a,(a^\star )^{:,-i}(\xi),\xi^i)=U^i_0((a^\star )^{:,i}(\xi),(a^\star )^{:,-i}(\xi),\xi^i).$$
\end{Definition}

Since the uniqueness of a Nash equilibrium is more the exception than the rule, we also need to assume that the community of Agents has agreed on a common rule to choose between different equilibria. More precisely, we are given a total order on $\R^N$, which we denote by $\succeq$. For, instance we could consider the order defined by, for any $(x,y)\in\R^N\times\R^N$
$$x\succeq y,\text{ if $\Sum_{i=1}^N\mathcal U^i(x^i)\leq \Sum_{i=1}^N\mathcal U^i(y^i)$},$$
for some given utility function $(\mathcal U^i)_{1\leq i\leq N}$, which means that the community of Agents prefers the Nash equilibria which maximize the total utility of the Agents. Moreover, we assume that if the community of Agents is indifferent (for the order $\succeq$) between several equilibriums, then it always chooses the ones which are the most profitable for the Principal.

\subsubsection{The best reaction functions}\label{sec:bestreac}
The first step towards finding potential Nash equilibria for the Agents is to be able to characterize the so-called reaction function of the Agents. We therefore start by solving the latter problem, for $i=1,\dots,N$, and for given contract $\xi$ and admissible actions of the other Agent $a^{-i}\in\mathcal M_{N,N-1}$. For the time being, we remain rather vague concerning the admissibility conditions for the contracts, since these will appear naturally later on. The following arguments can therefore be seen as heuristic in this respect. 

\vspace{0.5em}
Let us first define the value function of the an Agent $i$ by 
$$U_0^i(a^{-i},\xi^i):=\underset{a\in\mathcal A^i(a^{-i})}{\sup} \E^{\P^{a\otimes_ia^{-i}}}\left[-\exp\left(-R_A^i\left(\xi^i+\Gamma_i(X_T)-\int_0^Tk^i(s,a_s,X_s)ds\right)\right)\right].$$
The dynamic version of this stochastic control problem is then given, for any $t\in[0,T]$ by
$$U_t^i(a^{-i},\xi):=\underset{a\in\mathcal A^i(a^{-i})}{\esup} \E^{\P^{a\otimes_ia^{-i}}}\left[\left.-e^{-R_A^i\left(\xi^i+\Gamma_i(X_T)-\int_t^Tk^i(s,a_s,X_s)ds\right)}\right|\mathcal F_t\right].$$
Define then for any $a\in\mathcal A^i(a^{-i})$
$$U_t^i(a,a^{-i},\xi^i):= \E^{\P^{a\otimes_ia^{-i}}}\left[\left.-e^{-R_A^i\left(\xi^i+\Gamma_i(X_T)-\int_t^Tk^i(s,a_s,X_s)ds\right)}\right|\mathcal F_t\right].$$
Then, $e^{R_A^i\int_0^tk(s,a_s,X_s)ds}U_t^i(a,a^{-i},\xi)$ should be an $(\F,\P^{a\otimes_ia^{-i}})$-martingale. By the martingale representation property\footnote{To be perfectly rigorous, the representation property has to be applied for the measure $\P$, since there is no reason why in general $\P^a$ should still satisfy it. This means that one has to use Bayes formula to express $U_t^i(a,a^{-i},\xi)$ as a conditional expectation under $\P$ and then apply the representation property. We thank Sa\"id Hamad\`ene for pointing out that subtlety to us.}, there should therefore exist an $\R^N$- valued and $\F$-predictable process $\widetilde Z^{i,a,a^{-i},\xi^i}$ such that, after applying It\^o's formula
\begin{align*}
U_t^i(a,a^{-i},\xi^i)=&-e^{-R_A^i(\xi^i+\Gamma_i(X_T)}+\int_t^TR_A^iU_s^i(a,a^{-i},\xi^i)k^i(s,a_s,X_s)ds\\
&-\int_t^Te^{-R_A\int_0^sk^i(u,a_u,X_u)du}\widetilde Z^{i,a,a^{-i},\xi^i}_s\cdot \Sigma_s dW_s^{a\otimes_ia^{-i}},\ \qquad  0\leq t\leq T,\ a.s.
\end{align*}
By definition of $W^{a\otimes_ia^{-i}}$, we deduce that, for any $t\in[0,T]$,
\begin{align*}
U_t^i(a,a^{-i},\xi^i)=&-e^{-R_A^i(\xi^i+\Gamma_i(X_T)}+\int_t^TR_A^i U_s^i(a,a^{-i},\xi^i)Z^{i,a,a^{-i},\xi^i}_s\cdot \Sigma_s dW_s\\
&-\int_t^TR_A^iU_s^i(a,a^{-i},\xi^i)\left(b(s,a_s\otimes_ia^{-i}_s,X_s)\cdot Z^{i,a,a^{-i},\xi^i}_s- k^i(s,a_s,X_s)\right)ds,\ a.s.,
\end{align*}
where we have introduced the notation
$$Z^{i,a,a^{-i},\xi^i}_t:=-\frac{e^{-R_A^i\int_0^tk^i(s,a_s,X_s)ds}}{R_A^iU_t^i(a,a^{-i},\xi^i)}\widetilde Z^{i,a,a^{-i},\xi^i}_t,\ dt\times d\P-a.e.$$
Then, if we set 
$$Y^{i,a,a^{-i},\xi^i}_t:=-\frac{\log\left(-U_t^i(a,a^{-i},\xi^i)\right)}{R_A^i},\ t\in[0,T],\ a.s.,$$
we deduce by It\^o's formula (remember that $U^i(a,a^{-i},\xi^i)$ is positive by definition) that for any $t\in[0,T]$, $a.s.$,
\begin{align*}
Y^{i,a,a^{-i},\xi^i}_t=&\ \xi^i+\Gamma_i(X_T)-\int_t^TZ^{i,a,a^{-i},\xi^i}_s\cdot \Sigma_s dW_s\\
&+\int_t^T\left(-\frac{R_A^i}{2}\No{\Sigma_s Z^{i,a,a^{-i},\xi^i}_s}^2+b(s,a_s\otimes_ia^{-i}_s,X_s)\cdot Z^{i,a,a^{-i},\xi^i}_s- k^i(s,a_s,X_s)\right)ds.
\end{align*}
The above equation can be identified as a linear-quadratic backward SDE with terminal condition $\xi^i+\Gamma_i(X_T)$ and generator $\tilde f^{i,a^{-i}}:[0,T]\times\Omega\times\R^N\times\R^N\longrightarrow \R$, defined for any $(t,\omega,z,a)\in[0,T]\times\Omega\times\R^N\times\R^N$ by
$$\tilde f^{i,a^{-i}}(t,\omega,z,a):=-\frac{R_A^i}{2}\No{\Sigma_t z}^2+b(t,a\otimes_ia^{-i}_t(\omega),X_t(\omega))\cdot z- k^i(t,a,X_t(\omega)).$$

Define then, for any $(t,\omega,z)\in[0,T]\times\Omega\times\R^N$, the map $f^{i,a^{-i}}:[0,T]\times\Omega\times\R^N\longrightarrow \R$, by
$$ f^{i,a^{-i}}(t,\omega,z):=-\frac{R_A^i}{2}\No{\Sigma_t z}^2+\underset{a\in\mathcal A^i(a^{-i})}{\sup}\left\{b(t,a\otimes_ia^{-i}_t(\omega),X_t(\omega))\cdot z- k^i(t,a, X_t(\omega))\right\}.$$
Assume then that the BSDE with terminal condition $\xi^i+\Gamma_i(X_T)$ and generator $f^{i,a^{-i}}$ admits a maximal solution $(Y^{i,a^{-i},\xi^i},Z^{i,a^{-i},\xi^i})$, that is to say for any $t\in[0,T]$
\begin{equation}\label{eq:BSDE}
Y^{i,a^{-i},\xi^i}_t=\xi^i+\Gamma_i(X_T)+\int_t^Tf^{i,a^{-i}}\left(s,Z^{i,a^{-i},\xi^i}_s\right)ds-\int_t^TZ^{i,a^{-i},\xi^i}_s\cdot \Sigma_s dW_s,\ a.s.,
\end{equation}
and for any $(\tilde Y,\tilde Z)$ satisfying \reff{eq:BSDE}, we have $Y^{i,a^{-i},\xi^i}_\cdot\geq \tilde Y_\cdot,\ a.s.$

\vspace{0.5em}
Assume moreover that a comparison theorem holds for the maximal solution of \reff{eq:BSDE}. Then, since the supremum in the definition of $f^{i,a^{-i}}$ is always attained by Assumptions \ref{assump:b} and \ref{assump:k}, we deduce immediately that we actually have
$$Y^{i,a^{-i},\xi^i}_t=\underset{a\in\mathcal A^i(a^{-i})}{\esup}Y^{i,a,a^{-i},\xi^i}_t,\ t\in[0,T], \ a.s.,$$
which implies in turn that
$$U_t^i(a^{-i},\xi^i)=-\exp\left(-R_A^iY^{i,a^{-i},\xi^i}_t\right), \ t\in[0,T], \ a.s.,$$
and that the optimal action of Agent $i$, given a contract $\xi^i$ and actions $a^{-i}$ of the other Agents, is given by any 
$$a^{*,i,a^{-i},\xi^i}_t\in \underset{a\in\mathcal A^i(a^{-i})}{\rm argmax}\tilde f^{i,a^{-i}}\left(t,Z^{i,a^{-i},\xi^i}_t,a\right),\ dt\times d\P-a.e.$$
Of course, in order for our previous argumentation to be  truly meaningful, the BSDEs appearing above have to admit a maximal solution and to satisfy a comparison theorem. In order to discuss these questions, let us first establish the following lemma
\begin{Lemma}\label{lemma:fgrowth}
Let Assumptions \ref{assump:b} and \ref{assump:k} hold. Then, we have for some constant $C>0$
$$|f^{i,a^{-i}}(t,z)|\leq C\left(1+\No{a^{-i}}^2+\No{z}^{2}\right),$$
and any maximizer $a^\star $ in the definition of $f^{i,a^{-i}}$ satisfies
 $$\No{a^\star (t,z)}\leq C\left(1+\No{z}^\frac{1}{\ell -1}\right).$$
\end{Lemma}
\proof
First of all, notice that for any $(t,z)\in[0,T]\times\R^N$
\begin{align*}
f^{i,a^{-i}}(t,z):=&-\frac{R_A^i}{2}\No{\Sigma_t z}^2+\underset{a\in\mathcal A^i(a^{-i})}{\sup}\left\{\sum_{j=1}^Nb^j(t,(a\otimes_ia_t^{-i})^{:,j},X_t) z^j- k^i(t,a,X_t)\right\}.
\end{align*}

Then, we have
\begin{align*}
&\underset{a\in\mathcal A^i(a^{-i})}{\sup}\left\{\sum_{j=1}^Nb^j(t,(a\otimes_ia_t^{-i})^{:,j},X_t) z^j- k^i(t,a,X_t)\right\}\\
&\leq \underset{a\in\R^N}{\sup}\left\{\sum_{j=1}^Nb^j(t,(a\otimes_ia_t^{-i})^{:,j},X_t) z^j- k^i(t,a,X_t)\right\}\\
&=\sum_{j=1}^Nb^j\left(t,(a^\star (t,z,X_t)\otimes_ia_t^{-i})^{:,j},X_t\right) z^j- k^i\left(t,a^\star \left(t,z,X_t\right),X_t\right),
\end{align*}
where $a^\star (t,z,X_t)$ verifies the first-order conditions
$$\frac{\partial b^j}{\partial a^j}(t,(a^\star (t,z,X_t)\otimes_ia_t^{-i})^{:,j},X_t) z^j= \frac{\partial k^i}{\partial a^j}(t,a^\star (t,z,X_t),X_t),\ j=1,\dots,N.$$
Since $\nabla b^j$ and $\nabla k^i(t,a^\star (t,z,X_t),X_t)/\No{a^\star (t,z,X_t)}^{\ell-1}$ are bounded according to Assumptions \ref{assump:b} and \ref{assump:k}, we deduce immediately that, for some constant $C>0$
$$\No{a^\star (t,z,X_t)}\leq C\left(1+\No{z}^\frac{1}{\ell -1}\right).$$
Finally, since $\ell\geq 2$, we have $\ell/(\ell -1)\leq 2$, from which the desired result follows.
\ep

\vspace{0.5em}
Hence, by Lemma \ref{lemma:fgrowth}, the generator of the BSDE \reff{eq:BSDE} has quadratic growth in $z$. Thus, existence of a solution is ensured by the results of \cite{kob,bh,bek} for instance, as soon as $\xi^i\in M^\phi(\R)$. Whether a maximal solution exists, shall a priori require more assumptions. As for the comparison result, it is less clear, but since we will only use heuristically the result of this section, we do not try to address this question here (see nonetheless \cite{bh2,dhr,tev} for related results).

\subsubsection{Nash equilibria and multidimensional quadratic BSDEs}
The heuristic reasoning of the previous section naturally leads us to think that there should be a close connection between Nash equilibria between the $N$ Agents and the solutions (when they exist) of the following multidimensional BSDE
\begin{align}\label{eq:bsdemulti}
Y_t^\xi=\xi+\Gamma(X_T)+\int_t^Tf(s,Z^\xi_s,X_s)ds-\int_t^T(Z^\xi_s)^\top\Sigma_s dW_s,\ a.s.,
\end{align}
where the map $f:[0,T]\times\mathcal M_N(\R)\longrightarrow \R^N$ is defined, for $"=1,\dots,N$, by
$$f^i(t,z,x):=f^{i,(a^\star )^{:,-i}(s,z, x)}(s,z^{:,i}),\text{ for every $(s,z)\in[0,T]\times\mathcal M_N(\R)$},$$
with the matrix $a^\star (s,z,x)\in\mathcal M_N(\R)$ being defined, for $"=1,\dots,N$, and for every $(s,z,x)\in[0,T]\times\mathcal M_N(\R)\times\R^N$, by
\begin{equation}\label{eq:nash}
(a^\star )^{:,i}(s,z,x)\in \underset{a\in\mathcal A^i((a^\star )^{:,-i})}{\rm argmax}\left\{\sum_{j=1}^Nb^j(t,(a\otimes_i(a^\star _t)^{:,-i}(s,z,x))^{:,j},x) z^{i,j}- k^i(t,a,x)\right\},
\end{equation}
where it is implicitly assumed that a given maximizer has been chosen if there are more than one.
  
  \vspace{0.5em}
  The attentive reader should have realized that Equation \eqref{eq:nash} defining the map $a^\star$ was actually circular, since $a^\star$ appears on both sides of the equation. In general, it is not clear at all that $a^\star$ is then well-defined. We therefore need to impose an implicit assumption on the maps $b$ and $k$ as follows
  \begin{Assumption}\label{assump:nash}
  For every $(s,z,x)\in[0,T]\times\mathcal M_N(\R)\times\R^N$, there is at least one matrix $a^\star(s,z,x)$ in $\mathcal M_N(\R)$ such that for any $1\leq i\leq N$,
$$  (a^\star )^{:,i}(s,z,x)\in \underset{a\in\mathcal A^i((a^\star )^{:,-i})}{\rm argmax}\left\{\sum_{j=1}^Nb^j(t,(a\otimes_i(a^\star _t)^{:,-i}(s,z,x))^{:,j},x) z^{i,j}- k^i(t,a,x)\right\}.$$
We denote by $A^\star(s,z,x)$ the set of all matrices satisfying the above equation.
  \end{Assumption}
  A typical example where Assumption \ref{assump:nash} is satisfied, is when the map $b$ has a linear dependence with respect to the effort matrix $a$, since in this case the maximisation in \eqref{eq:nash} no longer depends on the other columns of $a^\star(s,z,x)$. For instance, one could consider the case
  $$b^i(t,a,x):=\tilde b^i(t,x)+a^i-\Sum_{j=1,\ j\neq i}^Na^j,$$
  which is nothing else than a more involved version of the benchmark case of Section \ref{sec:bench}.

  \vspace{0.5em}
  Let us now give a precise meaning to being a solution of \reff{eq:bsdemulti}. Let us first start by defining the following spaces. Fix some probability measure $\Q$ equivalent to $\P$ and a finite-dimensional normed vector space $E$, with a given norm $\No{\cdot}_E$
  
  \vspace{0.5em}
 ${\rm BMO}(\Q,\R^N)$ will denote the space of continuous square integrable $\F$-martingales $M$ (under $\Q$), $\R^N$ valued, such that $\No{M}_{{\rm BMO}(\Q)}<+\infty$, where
  $$\No{M}_{{\rm BMO}(\Q)}^2:=\underset{\tau\in\mathcal T_0^T}{\esup}\No{\mathbb E^{\Q}\left[\left.\Tr{\langle M\rangle_T}-\Tr{\langle M\rangle_{\tau}}\right|\mathcal F_\tau\right]}_{\infty}<+\infty,$$
where for any $t\in[0,T]$, $\mathcal T_t^T$ is the set of $\F$-stopping times taking their values in $[t,T]$ and where for any $p\in[1,+\infty],$ $\No{\cdot}_{p}$ denotes the usual norm on the space $L^p(\Omega,\mathcal F,\Q)$ of $\R$-valued random variables.

  \vspace{0.5em}
  $\H^2_{\rm BMO}(\Q,\mathcal M_N(\R))$ will denote the space of $\mathcal M_N(\R)$-valued, $\F$-predictable processes $Z$ such that $\No{Z}_{\H^2_{{\rm BMO}(\Q)}}<+\infty,$ where
  $$\No{Z}_{\H^2_{{\rm BMO}(\Q)}}:=\No{\int_0^TZ_sdW_s}_{{\rm BMO}(\Q)}.$$
  
   \vspace{0.5em}
$\H^2(\Q,E)$ will denote the space of $E$-valued, $\F$-predictable processes $Z$ s.t. $\No{Z}_{\H^2(\Q,E)}<+\infty,$ where
   $$\No{Z}_{\H^2(\Q,E)}:=\mathbb E^\Q\left[\int_0^T\No{ Z_s}_E^2ds\right]<+\infty.$$
   As usual, we denote by $\H^2_{\rm loc}(\Q,E)$ the localized version of this space. A solution of \reff{eq:bsdemulti} is then a pair $(Y^\xi,Z^\xi)$ such that $Y$ is a continuous $\F$-semimartingale satisfying \reff{eq:bsdemulti}, and $Z^\xi\in\H^2_{\rm loc}(\P,\mathcal M_N(\R))$. 

\vspace{0.5em}
Before stating the main result of this section, we need to introduce the so-called reverse H\"older inequality. 
\begin{Definition}[Reverse H\"older inequality]
Fix some probability measure $\Q$ equivalent to $\P$ and some $p>1$. A positive, or negative, $\F$-progressively measurable process $P$ is said to satisfy $R_p(\Q)$ if for some constant $C>0$
$$\underset{\tau\in\mathcal T_0^T}{\esup}\ \E^\Q\left[\left.\left(\frac{P_T}{P_\tau}\right)^p\right|\mathcal F_\tau\right]\leq C,\ a.s.$$
\end{Definition}

The link between existence of a Nash equilibrium between the Agents and existence of solutions to \reff{eq:bsdemulti} is given in the following theorem.

\begin{Theorem}\label{th:nash}
Let Assumptions \ref{assump:sigma}, \ref{assump:b}, \ref{assump:k} and \ref{assump:nash} hold. There is a one-to-one correspondence between
\begin{itemize}
\item[$(i)$] a Nash equilibrium $a^\star (\xi)\in\mathcal A$ such that for any $i=1,\dots,N$, there exists some $p>1$ such that
$$\left(\E^{\P^{a^\star (\xi)}}\left[\left.-e^{-R_A^i\left(\xi^i+\Gamma_i(X_T)-\int_0^Tk^i(s,(a^\star _s(\xi))^{:,i}, X_s)ds\right)}\right|\mathcal F_t\right]\right)_{t\in[0,T]}\text{ satisfies $R_p(\P^{a^\star (\xi)})$,}$$

\item[$(ii)$] a solution $(Y,Z)$ to \reff{eq:bsdemulti}, such that in addition $Z\in \H^2_{\rm BMO}(\P,\mathcal M_N(\R)),$
\end{itemize}
the correspondence being given by, for any $i=1,\dots,N$, $ds\times d\P-a.e.$,
$$(a^\star _s(\xi))^{:,i}\in \underset{a\in\mathcal A^i((a^\star )^{:,-i})}{\rm argmax}\left\{\sum_{j=1}^Nb^j(s,(a\otimes_i(a^\star _s)^{:,-i}(s,Z_s,X_s))^{:,j},X_s) Z_s^{i,j}- k^i(s,a,X_s)\right\}.$$
\end{Theorem}

\proof {\bf Step $1$.} We start by showing that $(i)$ leads to $(ii)$. For any $1\leq i\leq N$, and for any $\tau\in\mathcal T_{0,T}$, let us define the following family of random variables
$$U^i(\tau,\xi):=\underset{a\in\mathcal A^i((a^\star (\xi))^{:,-i})}{\esup} \E^{\P^{a\otimes_i(a^\star (\xi))^{:,-i}}}\left[\left.-e^{-R_A^i\left(\xi^i+\Gamma_i(X_T)-\int_\tau^Tk^i(s,a_s,X_s)ds\right)}\right|\mathcal F_\tau\right].$$
It is a classical result that this family satisfies the following dynamic programming principle (see for instance Theorem 2.4 in \cite{elkartan} as well as the discussion in their Section 2.4.2)
$$U^i(\tau,\xi)=\underset{a\in\mathcal A^i((a^\star (\xi))^{:,-i})}{\esup} \E^{\P^{a\otimes_i(a^\star (\xi))^{:,-i}}}\left[\left.e^{R_A^i\int_\tau^\theta k^i(s,a_s,X_s)}U^i(\theta,\xi)\right|\mathcal F_\tau\right],$$
for any $\theta\in\mathcal T_{0,T}$ such that $\tau\leq\theta,$ a.s.

\vspace{0.5em}
It is then immediate that for any $a\in\mathcal A^i((a^\star (\xi))^{:,-i})$, the family $(e^{R_A^i\int_0^\tau k^i(s,a_s,X_s)}U^i(\tau,\xi))_{\tau\in\mathcal T_{0,T}}$ is a so-called $\P^{a\otimes_i(a^\star (\xi))^{:,-i}}$-supermartingale system. Hence, by the results of \cite{delleng}, it can be aggregated by a unique (up to indistinguishability) $\F$-optional process, which actually coincides, a.s., with $(e^{R_A^i\int_0^t k^i(s,a_s,X_s)}U_t^i((a^\star (\xi))^{:,-i},\xi))_{t\in[0,T]}$ defined in the previous section. 

\vspace{0.5em}
Moreover, this aggregator remains a $\P^{a\otimes_i(a^\star (\xi))^{:,-i}}$-supermartingale, which then admits a c\`adl\`ag modification (remember that the filtration considered satisfies the usual conditions). Let us now check that $(e^{R_A^i\int_0^t k^i(s,(a^\star (\xi))^{:,i}_s,X_s)}U_t^i((a^\star (\xi))^{:,-i},\xi))_{t\in[0,T]}$ is a uniformly integrable $\P^{a^\star (\xi)}$-martingale.

\vspace{0.5em}
By definition of a Nash equilibria, $(a^\star (\xi))^{:,i}$ is optimal for the problem of Agent $i$, that is
$$U_0^i((a^\star (\xi))^{:,-i},\xi)=\E^{\P^{a^\star (\xi)}}\left[-e^{-R_A^i\left(\xi^i+\Gamma_i(X_T)-\int_0^Tk^i(s,(a^\star _s(\xi))^{:,i},X_s)ds\right)}\right].$$
Next, by the supermartingale property proved above (which holds for any choice of action of Agent $i$), and by definition of the value function of Agent $i$, we must have
\begin{align*}
U_0^i((a^\star (\xi))^{:,-i},\xi)&\geq \E^{\P^{a^\star (\xi)}}\left[e^{R_A^i\int_0^t k^i(s,(a^\star _s(\xi))^{:,i},X_s)}U_t^i((a^\star (\xi))^{:,-i},\xi)\right]\\
&\geq  \E^{\P^{a^\star (\xi)}}\left[-e^{-R_A^i\left(\xi^i+\Gamma_i(X_T)-\int_0^Tk^i(s,(a^\star _s(\xi))^{:,i},X_s)ds\right)}\right]\\
&=U_0^i((a^\star (\xi))^{:,-i},\xi),
\end{align*}
 for any $t\in[0,T]$. 
Hence, all these terms have to be equal, which implies in particular that
$$e^{R_A^i\int_0^t k^i(s,(a^\star _s(\xi))^{:,i},X_s)}U_t^i((a^\star (\xi))^{:,-i},\xi)=\E^{\P^{a^\star (\xi)}}\Big[-e^{-R_A^i(\xi^i+\Gamma_i(X_T)-\int_0^Tk^i(s,(a^\star _s(\xi))^{:,i},X_s)ds)}\Big|\mathcal F_t\Big],$$
 for any $t\in[0,T]$. This provides the desired result, since the right-hand side is obviously a $\P^{a^\star (\xi)}$-martingale, as the conditional expectation of an integrable random variable, and since by Assumptions \ref{assump:b} and \ref{assump:k} and by definition of admissible controls, this martingale actually belongs to $L^p(\Omega,\mathcal F,\P^{a^\star (\xi)})$, for any $p\geq 1$ (with moments uniformly bounded in $t\in[0,T]$ by Doob's inequality). Since it is also negative, by the predictable martingale representation property, there exists an $\F$-predictable process $ Z^{i,a^\star (\xi),\xi}\in \H^2(\P^{a^\star (\xi)},\R^N)$ such that, for any $t\in[0,T]$
\begin{align*}
e^{R_A^i\int_0^t k^i(s,(a^\star _s(\xi))^{:,i},X_s)}U_t^i((a^\star (\xi))^{:,-i},\xi)=U_0^i((a^\star (\xi))^{:,-i},\xi)\mathcal E\left(\int_0^\cdot Z_s^{i,a^\star (\xi),\xi}\cdot \Sigma_s dW_s^{a^\star (\xi)}\right)_t,\ a.s.
\end{align*}
Then, since by assumption $e^{R_A^i\int_0^t k^i(s,(a^\star _s(\xi))^{:,i}, X_s)}U_t^i((a^\star (\xi))^{:,-i},\xi)$ satisfies $(R_p(\P))$, we can use Theorems 3.3 and 3.4 of \cite{kaz} to deduce that $Z^{i,a^\star (\xi),\xi}$ belongs to both $\H^2_{\rm BMO}(\P^{a^\star (\xi)},\R)$ and $\H^2_{\rm BMO}(\P,\R)$. Furthermore, a simple application of It\^o's formula leads to
\begin{align*}
&U_t^i((a^\star (\xi))^{:,-i},\xi)\\
&=-e^{-R_A^i(\xi^i+\Gamma_i(X_T))}+\int_t^TR_A^i U_s^i((a^\star (\xi))^{:,-i},\xi) Z^{i,a^\star (\xi),\xi}_s\cdot \Sigma_s dW_s\\
&\hspace{0.9em}-\int_t^TR_A^iU_s^i((a^\star (\xi))^{:,-i},\xi)\left(b(s,a^\star _s(\xi),X_s)\cdot  Z^{i,a^\star (\xi),\xi}_s- k^i(s,(a^\star _s(\xi))^{:,i},X_s)\right)ds.
\end{align*}
Therefore, we deduce that for any $a\in\mathcal A^i((a^\star (\xi))^{:,-i})$, we have
\begin{align*}
&(R_A^i)^{-1}e^{R_A^i\int_0^t k^i(s,a_s)ds}U_t^i((a^\star (\xi))^{:,-i},\xi)\\
&=\frac{U_0^i((a^\star (\xi))^{:,-i},\xi)}{R_A^i}-\int_0^te^{R_A^i\int_0^s k^i(r,a_r,X_r)dr}U_s^i((a^\star (\xi))^{:,-i},\xi) Z^{i,a^\star (\xi),\xi}_s\cdot \Sigma_s dW^{a\otimes_i(a^\star (\xi))^{:,-i}}_s\\
&\hspace{0.7em}+\int_0^te^{R_A^i\int_0^s k^i(r,a_r,X_r)dr}U_s^i((a^\star (\xi))^{:,-i},\xi)(b(s,a^\star _s(\xi),X_s)\cdot  Z^{i,a^\star (\xi),\xi}_s- k^i(s,(a^\star _s(\xi))^{:,i},X_s))ds\\
&\hspace{0.7em}-\int_0^te^{R_A^i\int_0^s k^i(r,a_r,X_r)dr}U_s^i((a^\star (\xi))^{:,-i},\xi)(b(s,a_s\otimes_i(a^\star _s(\xi))^{:,-i},X_s)\cdot  Z^{i,a^\star (\xi),\xi}_s-k^i(s,a_s,X_s))ds.
\end{align*}
Now remember that this process must be a $\P^{a\otimes_i(a^\star (\xi))^{:,-i}}$-supermartingale. This therefore implies that we must have for any $a\in\mathcal A^i((a^\star (\xi))^{:,-i})$, $a.s.$,
$$b(s,a^\star _s(\xi),X_s)\cdot  Z^{i,a^\star (\xi),\xi}_s- k^i(s,(a^\star _s(\xi))^{:,i},X_s)\geq b(s,a_s\otimes_i(a^\star _s(\xi))^{:,-i},X_s)\cdot  Z^{i,a^\star (\xi),\xi}_s-k^i(s,a_s,X_s).$$
In other words
$$(a^\star _s(\xi))^{:,i}\in \underset{a\in\mathcal A^i((a^\star )^{:,-i})}{\rm argmax}\left\{b(s,a\otimes_i(a^\star _s(\xi))^{:,-i},X_s)\cdot Z^{i,a^\star (\xi),\xi}- k^i(s,a,X_s)\right\}.$$
Define then 
$$Y_t^i(a^\star (\xi),\xi):=-\frac{\log\left(-U_t^i((a^\star (\xi))^{:,-i},\xi)\right)}{R_A^i},\ \quad  t\in[0,T],\ a.s.$$
We have immediately for any $t\in[0,T]$, $a.s.$,
\begin{align*}
Y^{i}_t(a^\star (\xi),\xi)=&\ \xi^i+\Gamma_i(X_T)-\int_t^TZ^{i,a^\star (\xi),\xi}_s\cdot \Sigma_s dW_s\\
&+\int_t^T\left(-\frac{R_A^i}{2}\No{\Sigma_s Z^{i,a^\star (\xi),\xi}_s}^2+b(s,a^\star _s(\xi),X_s)\cdot Z^{i,a^\star (\xi),\xi}_s- k^i(s,(a_s^\star )^{:,i},X_s)\right)ds.
\end{align*}
Since all of this has to hold true for any $i=1,\dots,N$, we deduce that if we define $Y_t(a^\star (\xi),\xi)$ as the vector in $\R^N$ whose $i$th coordinate is $Y^{i}_t(a^\star (\xi),\xi)$, and if we define $Z^{a^\star (\xi),\xi}$ as the $N\times N$ matrix whose $i$th column is $ Z^{i,a^\star (\xi),\xi}$, then the pair $(Y_t(a^\star (\xi),\xi),Z^{a^\star (\xi),\xi})$ is a solution of the BSDE \reff{eq:bsdemulti}, such that $Z^{a^\star (\xi),\xi}\in  \H^2_{\rm BMO}(\P,\mathcal M_N(\R))$.

\vspace{0.5em}
{\bf Step $2$.} Conversely, let us be given a solution $(Y,Z)$ to \reff{eq:bsdemulti} s.t. $Z\in  \H^2_{\rm BMO}(\P,\mathcal M_N(\R))$. A classical measurable selection argument allows us to define a $\mathcal M_N(\R)$-valued $\F$-predictable process $a^\star $ such that for any $i=1,\dots,N$
$$(a^\star _s)^{:,i}\in \underset{a\in\mathcal A^i((a^\star )^{:,-i})}{\rm argmax}\left\{b(s,a\otimes_i(a^\star _s)^{:,-i}(Z_s,X_s),X_s)\cdot Z^{:,i}_s- k^i(s,a,X_s)\right\}.$$
Define then for any $i=1,\dots,N$
$$U^i_t:=-\exp\left(-R_A^iY_t^i\right),\ a.s.$$
We can then go backwards in all the computations of Step $1$, to obtain that, thanks to the fact that $Z\in \H^2_{\rm BMO}(\P,\mathcal M_N(\R))$, $(e^{R_A^i\int_0^t k^i(s,a_s,X_s)}U_t^i)_{t\in[0,T]}$ is a $\P^{a\otimes_i(a^\star )^{:,-i}}$-supermartingale for any $a\in\mathcal A^i((a^\star )^{:,-i})$, and that $(e^{R_A^i\int_0^t k^i(s,(a^\star _s)^{:,i},X_s)}U_t^i)_{t\in[0,T]}$ is a $\P^{a^\star }$-martingale. This uses in particular the fact that the Dol\'eans-Dade exponential of a BMO-martingale is a uniformly integrable martingale. By the martingale optimality principle, this implies that
$$U^i_0=U_0^i((a^\star )^{:,-i},\xi),$$
and that $(a^\star )^{:,i}$ is optimal for the problem of Agent $i$. Since this holds for any $i=1,\dots, N$, this means that $a^\star $ is a Nash equilibrium.

\vspace{0.5em}
Finally we have to check that the $a^\star $ we have defined is such that for any $i=1,\dots,N$, there exists some $p>1$ such that
$$\left(\E^{\P^{a^\star }}\left[\left.-e^{-R_A^i\left(\xi^i+\Gamma_i(X_T)-\int_0^Tk^i(s,(a^\star _s)^{:,i},X_s)ds\right)}\right|\mathcal F_t\right]\right)_{t\in[0,T]}\text{ satisfies $R_p(\P^{a^\star })$.}$$
But
$$-e^{-R_A^i\left(\xi^i+\Gamma_i(X_T)-\int_0^Tk^i(s,(a^\star _s)^{:,i},X_s)ds\right)}=e^{R_A^i\int_0^Tk^i(s,(a^\star _s)^{:,i},X_s)ds}U_T^i,$$
so that for any $t\in[0,T]$
\begin{align*}
X_t^i:=\E^{\P^{a^\star }}\left[\left.-e^{-R_A^i\left(\xi^i+\Gamma_i(X_T)-\int_0^Tk^i(s,(a^\star _s)^{:,i},X_s)ds\right)}\right|\mathcal F_t\right]&=e^{R_A^i\int_0^tk^i(s,(a^\star _s)^{:,i},X_s)ds}U_t^i\\
&=U_0^i\mathcal E\left(\int_0^\cdot Z^{i}_s\cdot \Sigma_s dW_s^{a^\star }\right)_t.
\end{align*}
Hence, since $Z\in  \H^2_{\rm BMO}(\P,\mathcal M_N(\R))$, we deduce from Theorem 3.4 in \cite{kaz} that the desired result holds.
\ep

\subsubsection{On existence of Nash equilibria and admissible contracts}
The main result of the previous section gives a complete characterization of the Nash equilibria for the Agents, which satisfy some integrability conditions, as solutions to the multidimensional quadratic BSDE \reff{eq:bsdemulti}. However, this does not address the question of existence of these equilibria, and this is exactly where the heart of the problem lies. Indeed, unlike in the one-dimensional case (that is $N=1$) wellposedness results for this kind of BSDE are extremely scarce in the literature. Tevzadze \cite{tev} was the first to obtain a wellposedness result in the case of a bounded and sufficiently small terminal condition. It was then proved by Frei and dos Reis \cite{fdr} and Frei \cite{f} that even in seemingly benign situations, existence of global solutions could fail. Later on, Cheredito and Nam \cite{cn}, Kardaras et al. \cite{khz}, Kramkov and Pulido \cite{kp,kp2}, Hu and Tang \cite{ht}, Jamneshan et al. \cite{jam}, or more recently Luo and Tangpi \cite{lt} all obtained some positive results, but only in particular instances, which do not readily apply in our setting. Recently, Xing and \v{Z}itkovi\'c \cite{xz} obtained quite general existence and uniqueness results, but in a Markovian framework.

\vspace{0.5em}
Therefore, since the Principal wants to offer contracts which reveal the actions of the Agents, he will never offer a contract for which a Nash equilibria between the Agents does not exist. For simplicity, let us introduce, for any $\xi\in\mathcal C^{FB}$, the set ${\rm NA}(\xi)$ of Nash equilibria associated to $\xi$ and which satisfy condition $(i)$ in Theorem \ref{th:nash} (which can be the empty set according to the above discussion). Furthermore, we also define
$${\rm NAI}(\xi):=\left\{a\in {\rm NA}(\xi),\ a\succeq b,\text{ for any $b\in {\rm NA}(\xi)$} \right\}.$$
Besides, we remind the reader that, in line with the classical literature on the subject, we assume that since the Agents are indifferent between Nash equilibria in ${\rm NAI}(\xi)$, the Principal can make them choose the one that suits him best.
 
 \vspace{0.5em}
This motivates the following definition of admissible contracts for the second-best problem
$$\mathcal C^{SB}:=\left\{\xi\in\mathcal C^{FB},\ {\rm NAI}(\xi)\text{ is non-empty}\right\}.$$
As a consequence of Theorem \ref{th:nash}, we know that for any $\xi\in\mathcal C^{SB}$, there exists a pair $(Y_0^\xi,Z^\xi)\in\R^N\times\H^2_{\rm BMO}(\P,\mathcal M_N(\R))$ such that
\begin{align}\label{eq:bsdemulti2}
\xi=Y_0^\xi-\Gamma(X_T)-\int_0^Tf(s,Z^\xi_s,X_s)ds+\int_0^T(Z^\xi_s)^\top\Sigma_s dW_s,\ a.s.
\end{align}
where, the optimal feedback control $a^\star$ being given by ???,  we recall that the vector function $f$ is defined by
$$f^i:(t,z,x)\mapsto-\frac{R_A^i}{2}\No{\Sigma_t z^{:,i}}^2-k^i(t,(a^{:,i})^\star (t,z,x),x)+\sum_{j=1}^Nb^j(t,(a^{:,j})^\star (t,z,x),x)z^{j,i}  \;.$$

\subsection{Solving the Principal problem}
\subsubsection{The general case}
For any $(Y_0,Z)\in\R^N\times\mathbb H^2_{\rm BMO}(\P,\mathcal M_N(\R))$, define
$$\xi^{Y_0,Z}:=Y_0-\Gamma(X_T)-\int_0^Tf(s,Z_s,X_s)ds+\int_0^T(Z_s)^\top\Sigma_s dW_s.$$

By \reff{eq:bsdemulti2}, we know that the set of admissible contracts $\mathcal C^{SB}$ is actually included in the set 
$$\left\{\xi^{Y_0,Z},\ (Y_0,Z)\in\R^N\times\mathbb H^2_{\rm BMO}(\P,\mathcal M_N(\R))\right\}.$$
As usual in Holmstr\"om-Milgrom type problems, we know that the value of the constant vector $Y_0$ in the contract will be fine-tuned so that each Agent receives exactly his reservation utility. But this exactly corresponds to choosing 
$$Y_0^i=L^i:=-\ln(-\underbar{U}^i)/R^i_A,\ 1\leq i\leq N.$$ We can therefore consider the following problem, which is, a priori, an upper bound of the Principal value function
\begin{align}
\label{eq:Principall}
\overline{U}_0^{P,SB}:=\underset{Z\in\mathbb H^2_{\rm BMO}(\P,\mathcal M_N(\R))}{\sup}\; \underset{a\in{\rm NAI}(\xi^{L,Z})}{\sup}\E^{\P^a}\left[-e^{-R_P(X_T-\xi^{L,Z})\cdot{\bf 1}_N}\right].
\end{align}

Then, under this form, we can interpret $\overline{\xi}^{L,Z}:=\xi^{L,Z}+\Gamma(X_T)$ as the terminal value of the following Markovian controlled diffusion
$$Y_t=L-\int_0^tf(s,Z_s,X_s)ds+\int_0^t(Z_s)^\top\Sigma_sdW_s,\ t\in[0,T],$$
where the control process is actually $Z$. Furthermore, for simplicity of notations, we assume that the maximizer in the definition of $f$ is actually unique. We are thus back into the realm of classical stochastic control, and, at least formally, we can identify $\overline{U}_0^{P,SB}$ with the value $v(0,0,L)$ of the unique\footnote{Of course uniqueness needs to be verified in practice} solution in an appropriate sense to the following HJB equation
\begin{align}\label{eq:HJB}
\begin{cases}
\left(v_t+G(\cdot,,v_x,v_y,v_{xx},v_{yy},v_{xy})\right)(t,x,y)=0,\ (t,x,y)\in[0,T)\times\R^{2N},\\
v(T,x,y)=-\exp\left(-R_P\left(x+\Gamma(x)-y\right)\cdot{\bf 1}_N\right),\ (x,y)\in\R^{2N},
\end{cases}
\end{align}
where
\begin{align*}
G(t,x,p,q,\gamma,\eta,\nu):=\underset{z\in\mathcal M_N(\R)}{\sup}&\left\{b(t,a^\star (t,z,x),x)\cdot p-f(t,z,x)\cdot q+\frac12{\rm Tr}\left[\Sigma_t\Sigma_t^\top\gamma\right]\right.\\
&\left.\hspace{0.9em}+\frac12{\rm Tr}\left[z^\top\Sigma_t\Sigma_t^\top z\eta\right]+{\rm Tr}\left[\Sigma_t\Sigma_t^\top z\nu\right]\right\}.
\end{align*}
The following result follows from a classical verification argument, see for instance \cite{fs}, so that we will not provide its proof.
\begin{Proposition}
Assume that the PDE \eqref{eq:HJB} admits a unique classical $($that is $C^{1,2,2})$ solution, and that the supremum in the definition of $G$ is attained for at least one $z^\star (t,x,p,q,\gamma,\eta,\nu)$. Assume that $z^\star _t\in\mathbb H^2_{\rm BMO}(\P,\mathcal M_N(\R)$ where
$$z^\star _t:=z^\star (t,X^\star _t,v_x(t,X^\star _t,Y^\star _t),v_y(t,X^\star _t,Y^\star _t),v_{xx}(t,X^\star _t,Y^\star _t),v_{yy}(t,X^\star _t,Y^\star _t),v_{xy}(t,X^\star _t,Y^\star _t)),$$
where $X^\star $ and $Y^\star $ are the unique solutions $($assumed to exist$)$ of the coupled SDEs
\begin{align*}
X^\star _t&=\int_0^tb(s,a^\star (s,z^\star _s,X^\star _s),X^\star _s)ds+\int_0^t\Sigma_sdW_s,\\
Y^\star _t&=Y_0-\int_0^tf(s,z^\star _s,X^\star _s)ds+\int_0^t(z^\star _s)^\top\Sigma_sdW_s.
\end{align*}
Assume furthermore that for any $Y_0\in\R^N$, the contract $\xi^{Y_0,z^\star }\in\mathcal C^{SB}$. Then, the value function of the Principal is given by $v(0,0,L)$.
\end{Proposition}
 Of course, as usual with verification type results, the above proposition is a bit disappointing. This is the reason why we consider in the next section a more specific problem for which the problem of the Principal can actually be solved diretly, without having to refer to the HJB equation \reff{eq:HJB}. This particular case consists in considering a linear comparison function $\Gamma$ as in Section \ref{sec:gammalinear}.

\subsubsection{The simpler setting of linear comparison map $\Gamma$}
We now focus on the particular case where $\gamma$ is linear and each agent compares the terminal value of his project to the average of those of all the projects. Namely, we again suppose as in Section \ref{sec:gammalinear} that Assumption \ref{assump:example} holds. The Principal problem can be written as
\begin{align}
\label{eq:Principal}
U_0^{P,SB}=\underset{\xi\in\mathcal C^{SB}}{\sup}\underset{a\in{\rm NAI}(\xi)}{\sup}\E^{\P^a}\left[-e^{-R_P(X_T-\xi)\cdot{\bf 1}_N}-\sum_{i=1}^N\rho_ie^{-R_A^i\left(\xi^i+\gamma_i\left(X^i_T-\overline X^{-i}_T\right)-\int_0^Tk^i(s,a_s^{:,i})ds\right)}\right],
\end{align}
where the Lagrange multipliers $(\rho_i)_{1\leq i\leq N}$ are once again here to ensure the participation constraints of the Agents. 
Plugging \reff{eq:bsdemulti2} in \reff{eq:Principal}, we deduce
\begin{align*}
U_0^{P,SB}=&\ \underset{\xi\in\mathcal C^{SB}}{\sup}\underset{a\in{\rm NAI}(\xi)}{\sup}\E^{\P^a}\Big[-\mathcal E\Big(\int_0^\cdot\Big({\bf 1}_N^\top Z_s^\xi+{\bf 1}_N^\top+\gamma^\top-(\overline\gamma^-)^\top\Big)\Sigma_s dW^a_s\Big)_Te^{R_PY_0^\xi\cdot{\bf1}_N}\\
&\times e^{-R_p\int_0^T\beta(s,Z_s^\xi)ds}-\sum_{i=1}^N\rho_ie^{-R_A^i(Y_0^\xi)^i}\mathcal E\Big(-R_A^i\int_0^\cdot((Z_s^\xi)^\top\Sigma_s dW^a_s)^i\Big)_T\Big],
\end{align*}
where the map $\beta:[0,T]\times\mathcal M_N(\R)\longrightarrow \R$ is defined by
\begin{align*}
\beta(t,z):=&\left({\bf{1}}_N+\gamma-\overline\gamma^-\right)\cdot b(t,a(z))-k(t,a(z))\cdot{\bf 1}_N-\sum_{i=1}^N\frac{R_A^i}{2}\No{\Sigma_t z^{:,i}}^2\\
&-\frac{R_P}{2}\No{\Sigma_t\left(z^\top{\bf 1}_N+{\bf 1}_N+\gamma-\overline\gamma^-\right)}^2.
\end{align*}
Now it is clear by Assumptions \ref{assump:b}, \ref{assump:k} and Lemma \ref{lemma:fgrowth} that the map $\beta$ is continuous in $z$ and goes to $-\infty$ as $\No{z}$ goes to $\infty$, so that it admits at least one deterministic maximizer, which we denote by $z^\star _t$. Then, recalling that $Z^\xi$ belongs to $\H^2_{\rm BMO}(\P,\mathcal M_N(\R))$ and thus also to $\H^2_{\rm BMO}(\P^a,\mathcal M_N(\R))$ for any $a\in\mathcal A$ by Theorem 3.3 in \cite{kaz}, we deduce that
$$U_0^{P,SB}\leq \underset{Y_0^\xi\in\R^N}{\sup}\left\{-e^{R_PY_0^\xi\cdot{\bf1}_N-\int_0^T\beta(s,z^\star _s)ds}-\sum_{i=1}^N\rho_ie^{-R_A^i(Y_0^\xi)^i}\right\}.$$
The right-hand side is a concave function of $Y_0^\xi$ on $\R^N$, whose supremum is attained at $Y_0^\star $, with
\begin{align*}
(Y_0^\star )^i:=&\ \frac{1}{R_A^i}\log\left(\frac{\rho_iR_A^i}{R_P}\right)-\frac{R_P\overline R_A}{R_A^i\left(\overline R_A+NR_P\right)}\sum_{j=1}^N\frac{1}{R_A^j}\log\left(\frac{\rho_jR_A^j}{R_P}\right)\\
&+\frac{\overline R_A}{R_A^i\left(\overline R_A+NR_P\right)}\int_0^T\beta(s,z^\star _s)ds, \ i=1,\dots,N,
\end{align*}
and we directly deduce that
\begin{equation}\label{eq:secbest}
U_0^{P,SB}\leq -\frac{\overline R_A+NR_P}{\overline R_A}e^{\frac{\overline R_A}{\overline R_A+NR_P}\left(\sum_{i=1}^N\frac{1}{R_A^i}\log\left(\frac{\rho_iR_A^i}{R_P}\right)-\int_0^T\beta(s,z^\star _s)ds\right)}.
\end{equation}

In order to attain this upper bound, we would like to consider the contract $\xi^\star $ defined by
$$\xi^\star :=Y_0^\star -{\rm diag}(\gamma)(X_T-\overline X_T^-)-\int_0^Tf(s,z^\star _s)ds+\int_0^Tz^\star _s\Sigma_s dW_s.$$
But under this form, it is clear that the corresponding BSDE \reff{eq:bsdemulti} admits at least as a solution $(Y^\star ,z^\star )$, where $$Y^\star _t:=Y_0^\star -\int_0^tf(s,z^\star _s)ds+\int_0^t(z^\star _s)^\top\Sigma_s dW_s,\ t\in[0,T],\ a.s.$$
Moreover, since $z^\star $ is deterministic, it is clearly in $\H^2_{\rm BMO}(\P,\mathcal M_N(\R))$. Hence, we do have $\xi^\star \in\mathcal C^{SB}$, so that equality holds in \reff{eq:secbest}.

\vspace{0.5em}
Finally, it remains to choose the Lagrange multipliers so as to satisfy the participation constraints of the Agents, with reservation utilities $(\underbar{U}^i)_{1\leq i\leq N}$. We must therefore solve
$$-e^{-R_A^i(Y^\star _0)^i}=\underbar{U}^i,\ i=1,\dots, N.$$
Arguing exactly as in the first-best case, we compute that
$$\rho_i=-\frac{R_P}{\underbar{U}^iR_A^i}\left(\Prod_{j=1}^N(-\underbar{U}^j)^{\frac{R_P}{R_A^j}}\right)^{-1}e^{-\int_0^T\beta(s,z^\star _s)ds}.$$
We have thus proved
\begin{Theorem}\label{th:secbest}
Let Assumptions \ref{assump:sigma}, \ref{assump:b}, \ref{assump:k}, \ref{assump:example} and \ref{assump:nash} hold. Then an optimal contract $\xi_{SB}\in\mathcal C^{SB}$ in the problem \reff{eq:secbest}, with reservation utilities $(\underbar{U}^i)_{1\leq i\leq N}\in(-\infty,0)^N$, is given, for $i=1,\dots,N$, by
\begin{align*}
\xi_{SB}^i:=& -\frac{1}{R_A^i}\log(-\underbar{U}^i)-\gamma_i(X_T^i-\overline X_T^{-i})+\left(\int_0^Tz_s^\star dX_s\right)^i+\int_0^Tk^i(s,(a^\star _s((z^\star _s)^{i,:}))^{:,i})ds \\
&+\frac{R_A^i}{2}\int_0^T\No{\Sigma (z^\star _s)^{:,i}}^2ds,
\end{align*}
where the matrix $a^\star _s(z)\in\mathcal M_N(\R)$ is defined implicitly , for every $(s,z)\in[0,T]\times\mathcal M_N(\R)$ and for $"=1,\dots,N$, by 
$$(a^\star _s(z))^{:,i}\in \underset{a\in\mathcal A^i((a^\star )^{:,-i})}{\rm argmax}\left\{b(t,a\otimes_i(a^\star _s(z))^{:,-i})\cdot z^{:,i}- k^i(t,a)\right\},$$
and where $z^\star _t$ is any maximizer of the map
\begin{align*}
z\longmapsto&\left({\bf{1}}_N+\gamma-\overline\gamma^-\right)\cdot b(t,a^\star _t(z))-k(t,a^\star _t(z))\cdot{\bf 1}_N-\sum_{i=1}^N\frac{R_A^i}{2}\No{\Sigma_t z^{:,i}}^2\\
&-\frac{R_P}{2}\No{\Sigma_t\left(z^\top{\bf 1}_N+{\bf 1}_N+\gamma-\overline\gamma^-\right)}^2.
\end{align*}
\end{Theorem}
\subsubsection{Back to the bidimensional linear-quadratic  benchmark setting}
We once again focus on the benchmark case developed in Section \ref{sec:bench} for which explicit computations are available. In this framework, let's recall that 
$$b(t,a):=\begin{pmatrix}
a^{1,1}-a^{1,2}\\
a^{2,2}-a^{2,1}
\end{pmatrix},\ \text{for any $a:=\begin{pmatrix}a^{1,1} & a^{1,2}\\
a^{2,1} & a^{2,2}\end{pmatrix}\in\mathcal M_2(\R),$}$$
and for some constants $(k^{1,1},k^{1,2},k^{2,1},k^{2,2})\in(\R_+^\star )^4$
$$k(t,a):=\begin{pmatrix}
\frac{k^{1,1}}{2}\abs{a^{1,1}}^2+\frac{k^{2,1}}{2}\abs{a^{2,1}}^2\\
\frac{k^{2,2}}{2}\abs{a^{2,2}}^2+\frac{k^{1,2}}{2}\abs{a^{1,2}}^2
\end{pmatrix},\ \text{for any $a:=\begin{pmatrix}a^{1,1} & a^{1,2}\\
a^{2,1} & a^{2,2}\end{pmatrix}\in\mathcal M_2(\R).$}$$

Easy computations show that the optimal control matrix $a^\star (z)$ is given by
$$a^\star (z)=\begin{pmatrix}
\frac{z^{1,1}}{k^{1,1}} &\frac{z^{1,2}}{k^{1,2}}\\
\frac{z^{2,1}}{k^{2,1}} &\frac{z^{2,2}}{k^{2,2}}
\end{pmatrix}.$$
To find the optimal value $z^\star $, we therefore need to maximize the map
\begin{align*}
g:z\mapsto &\ (1+\gamma_1-\gamma_2)\left(\frac{z^{1,1}}{k^{1,1}}-\frac{z^{1,2}}{k^{1,2}}\right)+(1+\gamma_2-\gamma_1)\left(\frac{z^{2,2}}{k^{2,2}}-\frac{z^{2,1}}{k^{2,1}}\right)-\frac{\abs{z^{1,1}}^2}{2k^{1,1}}\\
&-\frac{\abs{z^{1,2}}^2}{2k^{1,2}}-\frac{\abs{z^{2,1}}^2}{2k^{2,1}}-\frac{\abs{z^{2,2}}^2}{2k^{2,2}}-\frac{\sigma_1^2}{2}\left(R_A^1\abs{z^{1,1}}^2+R_A^2\abs{z^{1,2}}^2\right)\\
&-\frac{\sigma_2^2}{2}\left(R_A^1\abs{z^{2,1}}^2+R_A^2\abs{z^{2,2}}^2\right)-\frac{R_P}{2}\sigma_1^2\left(z^{1,1}+z^{2,1}+1+\gamma_1-\gamma_2\right)^2\\
&-\frac{R_P}{2}\sigma_2^2\left(z^{1,2}+z^{2,2}+1+\gamma_2-\gamma_1\right)^2.
\end{align*}
Easy (but lengthy) calculations show that $g$ is actually concave, and admits a unique critical point given by, for $i,j=1,2$
\begin{align*}
(z^\star)^{i,i}=&\ \frac{1}{\alpha^{i,j}}
\left( (1+k^{i,i}k^{i,j}R_pR_A^i\sigma_1^2\sigma_2^2)(1+\gamma_i-\gamma_j) + 2 k^{i,i}k^{i,j}\left|R_p\sigma_i^2\right|^2 \right)\\
(z^\star)^{j,i}=&\ - \frac{1}{\alpha^{i,j}},
\left( (2+\sigma_i^2k^{i,i}R^i_A)(1+\gamma_i-\gamma_j) + R_p\sigma_i^2 k^{i,j} ( 1+\sigma_i^2 k^{i,i}(R_A^i+R_p))(1+\gamma_j-\gamma_i) \right),
\end{align*}
where
$$\alpha^{i,j}:=1+\sigma_i^2(R_A^i+R_P)k^{i,i}+(\sigma_j^2R_A^i+\sigma_i^2R_P)k^{j,i}+\sigma_i^2R_A^i(\sigma_j^2(R_A^i+R_P)+\sigma_i^2R_P)k^{j,i}k^{i,i}.$$

 Whenever $\gamma_1$ is significantly higher than $\gamma_2$, observe that once again  Agent $2$ optimal strategy can be to work against its own project. 

\subsubsection{Economic interpretation}

As in the first best case developed above, observe that the optimal contract is linear. Moreover each agent  obtains his reservation utility and is paid a given proportion of the value of each project. As observed explicitly in the linear-quadratic benchmark case, the analytic form of the optimal contract is more intricate than in the first best setting, but most qualitative properties of the optimal contract and strategies remain unchanged. In particular, each agent has incentives to help the project of a very competitive colleague, whereas he may have  to work against the project of a poorly competitive one. In some sense, a competitive agent rewards himself via his appetite for competition, and therefore it is in the economic interest of the Principal to provide him a project with higher probability of success, e.g. with less volatility or the help of the other Agents.

\vspace{0.5em}
We recall that  we assumed throughout the paper that the competition indexes $\gamma$ of candidates are observable for the Principal. In such framework, we emphasize that our results imply in particular that it is not optimal for the Principal to hire Agents with similar appetence for competition. Economic benefits for the firm (i.e. the Principal) follows from a diversification of competitive profiles, while each  Agent always recovers his reservation utility. A pertinent future research topic may be  to inquire if such conclusions remain valid, in an adverse selection  framework, where the competitive appetence of the agent are ex-ante unknown, and shall be determined via the design of a well suited menu of contracts.


\begin{thebibliography}{aa12}
\small 


 \bibitem{bek} Barrieu, P., El Karoui, N. (2013). Monotone stability of quadratic
semimartingales with applications to unbounded general quadratic BSDEs, {\sl The Annals of Probability}, 41(3B):1831--1863.

\bibitem{barts1}
Bartling, B., von Siemens, F.A. (2004). Inequity aversion and moral hazard with multiple agents, working paper, \url{http://citeseerx.ist.psu.edu/viewdoc/download?doi=10.1.1.408.646&rep=rep1&type=pdf}.

\bibitem{barts2}
Bartling, B., von Siemens, F.A. (2010). The intensity of incentives in firms and markets: moral hazard with envious agents, {\sl Labour Economics}, 17:598--607.

\bibitem{BD} Bolton, P., Dewatripont, M. (2005). Contract theory, {\sl MIT Press}, Cambridge.

\bibitem{borch}
Borch, K. (1962). Equilibrium in a reinsurance market, {\sl Econometrica}, 30:424--444.

\bibitem{bh} Briand, Ph., Hu, Y. (2006).
BSDE with quadratic growth and unbounded terminal value, {\sl Probability Theory and Related Fields}, 136(4):604--618.

\bibitem{bh2} Briand, Ph., Hu, Y. (2008).
Quadratic BSDEs with convex generators and unbounded terminal conditions, {\sl Probability Theory and Related Fields}, 141(3-4):543--567.

\bibitem{CCZ}Cadenillas, A., Cvitani\'c, J., Zapatero, F. (2007). Optimal risk-sharing
with effort and project choice, {\it Journal of Economic Theory}, 133:403--440.

%

\bibitem{cn}
Cheredito, N., Nam, K. (2015). Multidimensional quadratic and subquadratic BSDEs with special structure, {\sl Stochastics An International Journal of Probability and Stochastic Processes}, to appear, {\sl arXiv:1309.6716}.

\bibitem{cpt}
Cvitani\'c, J., Possama\"i, D., Touzi, N. (2014). Moral hazard in dynamic risk management, {\sl Management Science}, to appear, {\sl arXiv:1406.5852.}

\bibitem{cpt2}
Cvitani\'c, J., Possama\"i, D., Touzi, N. (2015). Dynamic programming approach to principal-agent problems, preprint, {\sl arXiv:1510.07111.}


\bibitem{CWZ5} Cvitani\'c J., Wan X., Zhang (2009). Optimal compensation with hidden action and lump-sum payment in a continuous-time model, {\it Applied Mathematics and Optimization}, 59:99--146.

\bibitem{CZ2}
Cvitani\'c, J., Wan, X., Zhang, J. (2006). Optimal contracts in continuous-time models, {\sl Journal of Applied Mathematics and Stochastic Analysis}, 1--27.


\bibitem{CvitanicZhang}
Cvitani\'c, J., Zhang, J. (2012). Contract theory in continuous time models, {Springer-Verlag}.

\bibitem{dhr}
Delbaen, F., Hu, Y., Richou, A. (2011). On the uniqueness of solutions to quadratic BSDEs with convex generators and unbounded terminal conditions, {\sl Annales de l'Institut Henri Poincar\'e - Probabilit\'es et Statistiques}, 47(2):559--574.

\bibitem{delleng}
Dellacherie, C., Lenglart, E. (1981). Sur des probl\`emes de r\'egularisation, de recollement et d'interpolation en th\'eorie des martingales, {\sl S\'eminaire de probabilit\'es $($Strasbourg$)$}, 15:328--346.


\bibitem{demou1}
Demougin, D., Fluet, C. (2003). Inequity aversion in tournaments, {\sl Working Paper 03-22, CIRP\'EE}, \url{http://papers.ssrn.com/sol3/papers.cfm?abstract_id=409560}. 

\bibitem{demou2}
Demougin, D., Fluet, C. (2006). Group vs. individual performance pay when workers are envious, {\sl In D. De-
mougin and C. Schade, eds., Contributions to Entrepreneurship and Economics - First Haniel-Kreis Meeting on Entrepreneurial Research}, Duncker \& Humbolt, 39--47.

\bibitem{demou3}
Demougin, D., Fluet, C., Helm, C. (2006). Output and wages with inequality averse agents, {\sl Canadian Journal of Economics} 39:399--413.

\bibitem{ds}
Demski, J.S., Sappington, D. (1984). Optimal incentives contracts with multiple agents, {\sl Journal of Economic Theory}, 33:152--171.

\bibitem{dje1}
Djehiche, B., Helgesson, P. (2014). The principal-agent problem; a stochastic maximum principle approach, preprint, {\sl arXiv:1410.6392}.

\bibitem{dje2}
Djehiche, B., Helgesson, P. (2015). The principal-agent problem with time inconsistent utility functions, preprint, {\sl arXiv:1503.05416}.

\bibitem{duffie}
Duffie, D., Geoffard, P.Y., Skiadas, C. (1994). Efficient and equilibrium allocations with stochastic differential utility, {\sl Journal of Mathematical Economics}, 23(2):133--146.

\bibitem{dg}
Dur, R., Glazer, A. (2004). Optimal incentive contracts when workers envy their boss, {\sl Tinbergen Institute Discussion Paper No. TI 2004-046/1}, \url{http://papers.ssrn.com/sol3/papers.cfm?abstract_id=536622}. 


\bibitem{elkarq}
	El Karoui, N., Quenez, M.-C. (1995). Dynamic programming and pricing of contingent claims in an incomplete market, {\sl SIAM Journal on Control and Optimization}, 33(1):27--66.
	
\bibitem{ekpq}
	El Karoui, N., Peng, S., Quenez, M.-C. (1997). 
	Backward stochastic differential equations in finance, {\sl Mathematical Finance}, 7(1):1--71.
	
\bibitem{elkartan}
El Karoui, N., Tan, X. (2013). Capacities, measurable selection and dynamic programming part II: application in stochastic control problems, preprint, {\sl arXiv:1310.3364}.

\bibitem{ew}
Englmaier, F., Wambach, A. (2002). Contracts and inequity aversion, {\sl CESifo working paper No. 809}, \url{http://papers.ssrn.com/sol3/papers.cfm?abstract_id=358325}.

\bibitem{ew2}
Englmaier, F., Wambach, A. (2010). Optimal incentive contracts under inequity aversion, {\sl Games and Economic Behaviour}, 69(2):312--328. 

\bibitem{et}
Espinosa, G.-E., Touzi, N. (2015). Optimal investment under relative performance concerns, {\sl Mathematical Finance}, 25(2):221--257.

\bibitem{fehr}
Fehr, E., Schmidt, K.M. (1999). A theory of fairness, competition, and cooperation, {\sl Quarterly Journal of Economics}, 114:817--868.

\bibitem{fehr2}
Fehr, E., Klein, A., Schmidt, K.M. (2001). Fairness, incentives, and contractual incompleteness, {\sl CEPR Discussion Paper No. 2790}, \url{http://papers.ssrn.com/sol3/papers.cfm?abstract_id=269389}.
\bibitem{fs}
Fleming, W.H., Soner, H.M. (1993). Controlled Markov processes and viscosity solutions, {\sl Applications of Mathematics 25}, Springer-Verlag, New York.

\bibitem{f}
Frei, C. (2014). Splitting multidimensional BSDEs and finding local equilibria, {\sl Stochastic Processes and their Applications}, 124(8):2654--2671.

\bibitem{fdr}
Frei, C., dos Reis, G. (2011). A financial market with interacting investors: does an equilibrium exist?, {\sl Mathematics and Financial Economics}, 4(3):161--182.

%

\bibitem{gw}
Goukasian, L., Wan, X. (2010). Optimal incentive contracts under relative income concerns, {\sl Mathematics and Financial Economics}, 4:57--86.
 
\bibitem{gs}
Green, J., Stokey, N. (1983). A comparison of tournaments and contracts, {\sl The Journal of Political Economy}, 91:349--364.

\bibitem{gross}
Grossman, S., Hart, O. (1983). An analysis of the principal-agent problem, {\sl Econometrica}, 51:7--45.

\bibitem{grund}
Grund, C., Sliwka, D. (2005). Envy and compassion in tournaments, {\sl Journal of Economics \& Management Strategy}, 14(1):187--207.


\bibitem{harr}
Harris, M., Kriebel, C., Raviv, A. (1982). Asymmetric information, incentives, and intrafirm resource allocation, {\sl Management Science}, 28:604--620.

\bibitem{HS8} Hellwig M., Schmidt, K.M. (2002). Discrete-time approximations of Holmstr\"om-Milgrom Brow-nian-motion model of intertemporal Incentive provision, {\em Econometrica}, 70:2225--2264.

\bibitem{holm}
Holmstr\"om, B. (1979). Moral hazard and observability, {\sl Bell Journal of Economics}, 10:74--91.

\bibitem{holmm}
Holmstr\"om, B. (1982). Moral hazard in teams, {\sl Bell Journal of Economics}, 13:324--340.

\bibitem{hm} Holmstr\"om B., Milgrom, P. (1987). Aggregation and linearity in the
provision of intertemporal incentives, {\em Econometrica}, 55(2):303--328.

\bibitem{ht}
Hu, Y., Tang, S. (2014). Multi-dimensional backward stochastic differential equations of diagonally quadratic generators, preprint, {\sl arXiv:1408.4579}.

\bibitem{itoh}
Itoh, H. (2004). Moral hazard and other-regarding preferences, {\sl The Japanese Economic Review}, 55(1):18--45.
 
\bibitem{jam}
Jamneshan, A., Kupper, M., Luo, P. (2015). Multidimensional quadratic BSDEs with separated generators, preprint, {\sl arXiv:1501.0046}.

\bibitem{jew}
Jewitt, I. (1988). Justifying the first-order approach to principal-agent problems, {\sl Econometrica}, 56:1177--1190.

\bibitem{khz}
Kardaras, C., Xing, H., \v{Z}itkovi\'c, G. (2015). Incomplete stochastic equilibria with exponential utilities close to Pareto optimality, preprint, {\sl arXiv:1505.07224}.

\bibitem{kaz}
Kazamaki, N. (1994). Continuous exponential martingales and BMO, {\sl Lecture Notes in Mathematics, vol. 1579}, Springer, Berlin.

\bibitem{kob}
Kobylanski, M. (2000). Backward stochastic differential equations and partial differential equations with quadratic growth, {\sl The Annals of Probability}, 28(2):259--276.

\bibitem{koo}
Koo, H.K., Shim, G., Sung, J. (2008). Optimal multi-Agent performance measures for team contracts, {\sl Mathematical Finance}, 18(4):649--667.

\bibitem{kragl}
Kragl, J. (2015). Group versus individual performance pay in relational employment contracts when workers are envious, {\sl Journal of Economics \& Management Strategy}, 24(1):131--150.

\bibitem{kp}
Kramkov, P., Pulido, S. (2014). A system of quadratic BSDEs arising in a price impact model, {\sl The Annals of Applied Probability}, to appear, {\sl arXiv:1408.0916}.

\bibitem{kp2}
Kramkov, P., Pulido, S. (2014). Stability and analytic expansions of local solutions of systems of quadratic BSDEs with applications to a price impact model, preprint, {\sl arXiv:1410.6144}.

   \bibitem{laffont}
Laffont, J.J., Martimort, D. (2001). The theory of incentives: the principal-agent model, {\sl Princeton
University Press,} Princeton.
   

\bibitem{lt}
Luo, P., Tangpi, L. (2015). Solvability of coupled FBSDEs with quadratic and superquadratic growth, preprint, {\sl arXiv:1505.01796}.



 
 \bibitem{mir1}
 Mirrlees, J.A. (1972). Population policy and the taxation of family size, {\sl Journal of Public Economics,} 1:169--198.
 
 
 \bibitem{mir3}
 Mirrlees, J.A. (1976). The optimal structure of incentives and authority within an organisation, {\sl Bell Journal of Economics}, 7:105--131.
 
 \bibitem{mir4}
 Mirrlees, J.A. (1999). The theory of moral hazard and unobservable behaviour: part I, {\sl Review of Economic Studies}, 66:3--21 (reprint of the 1975 unpublished version).
 
  \bibitem{mook}
  Mookherjee, D. (1984). Optimal incentive schemes with many agents, {\sl Review of Economic Studies}, 51:433--446.
  
 \bibitem{M1}  M\"uller, H. (1998). The first-best sharing rule in the
continuous-time Principal Agent problem with exponential utility.
{\sl Journal of Economic Theory} 79:276--280.

\bibitem{Mul2} M\"uller H. (2000). Asymptotic efficiency in dynamic principal-agent problems, {\em Journal of Economic Theory}, 91:292--301.

\bibitem{ns}
Nalebuff, B., Stiglitz, J. (1983). Prices and incentives: towards a general theory of compensation and competition, {\sl Bell Journal of Economics}, 14:21--43.

\bibitem{neilson1}
Neilson, W.S., Stowe, J. (2004). Incentive pay for other-regarding workers, working paper, \url{https://faculty.fuqua.duke.edu/~stowe/bio/Professional\%20Page/incentivepay.pdf}

\bibitem{neilson2}
Neilson, W.S., Stowe, J. (2008). Piece-rate contracts for other-regarding workers, {\sl Economic Inquiry}, 48(3):575--586.

   \bibitem{ou} Ou-Yang, H. (2003). Optimal contracts in a continuous-time
delegated portfolio management problem, {\it The Review of Financial
Studies}, 16:173--208.

\bibitem{ou2}
Ou-Yang, H. (2005). An equilibrium model of asset pricing and moral hazard, {\sl The Review of Financial Studies} 18:1219--1251.




\bibitem{rao1}
Rao, M.M., Ren, Z.D. (1991). Theory of Orlicz spaces, {\sl Marcel Dekker}, New York.
   
\bibitem{rao2}
Rao, M.M., Ren, Z.D. (2002). Applications of Orlicz spaces, {\sl Marcel Dekker,} New York, Basel. 
   
   \bibitem{rey}
   Rey-Biel, P. (2008). Inequity aversion and team incentives, {\sl Scandinavian Journal of Economics}, 110(2):297--320.
   
   \bibitem{roger}
   Rogerson, W.P. (1985). The first-order approach to Principal-Agent problems, {\sl Econometrica}, 53(6):1357--1368.
   
   
\bibitem{ross}
Ross, S.A. (1973). The economic theory of agency: the Principal's problem, {\sl Am. Econ. Rev.}, 63:134--139. Papers and proceedings of the eighty-fifth Annual Meeting of the American Economic Association.

\bibitem{sala} Salani\'e, B. (2005). The economics of contracts: a primer, {\sl MIT Press}, Cambridge.


\bibitem{sann} Sannikov, Y. (2008). A continuous-time version of the principal-agent
problem, {\it The Review of Economic Studies}, 75:957--984.

\bibitem{sann2}
Sannikov, Y. (2012). Contracts: the theory of dynamic principal-agent relationships and the continuous-time approach, working paper, Princeton University, \url{http://www.princeton.edu/~sannikov/congress_paper.pdf}.

\bibitem{SS3} Sch\"{a}ttler H., Sung, J. (1993). The first-order approach to
continuous-time principal-agent problem with exponential utility, {\em Journal of Economic Theory}, 61:331--371.

\bibitem{SS7} Sch\"attler H., Sung, J. (1997). On optimal sharing rules in
discrete- and continuous-time principal-agent problems with
exponential utility, {\em Journal of Economic Dynamics and Control}, 21:551--574.

	
	\bibitem{szec}
	Spence, M., Zeckhauser, R. (1971). Insurance, information, and individual action, {\sl American Economic Review}, 61:381--387.
	
  \bibitem{Su}
Sung, J. (1995). Linearity with project selection and controllable diffusion rate
in continuous-time principal-agent problems, {\sl The RAND Journal of Economics} 26:720--743.

\bibitem{S2} Sung J. (1997). Corporate insurance and managerial incentives, {\em Journal
of Economic Theory}, 74:297--332.

\bibitem{S3}
Sung, J. (2001). Lectures on the theory of contracts in corporate finance: from discrete-time to continuous-time models, {\sl Com2Mac Lecture Note Series, vol. 4.}, Pohang University of Science and Technology, Pohang, \url{http://down.cenet.org.cn/upfile/61/20081126192246136.pdf}.


\bibitem{tev}
Tevzadze, R. (2008). Solvability of backward stochastic differential equations with quadratic growth, {\sl Stochastic Processes and their Applications}, 118(3):503--515.

\bibitem{Williams} Williams N. (2009). On dynamic principal-agent problems in continuous
time, working paper, University of Wisconsin, \url{http://www.ssc.wisc.edu/~nwilliam/dynamic-pa1.pdf}.

\bibitem{wilson}
Wilson, R. (1968). The theory of syndicates, {\sl Econometrica}, 36:119--132.

\bibitem{xz}
Xing, H., \v{Z}itkovi\'c, G. (2016). A class of globally solvable Markovian quadratic BSDE systems and applications, preprint, {\sl arXiv:1603.00217}.

\bibitem{zec}
Zeckhauser, R. (1970). Medical insurance: a case study of the tradeoff between risk spreading and appropriate incentives, {\sl Journal of Economic Theory}, 2:10--26.


\end{thebibliography}
\end{document}